\documentclass[twocolumn,showpacs]{revtex4-1}
\usepackage{graphicx}
\usepackage{dcolumn}
\usepackage{bm}
\usepackage{bbm}
\usepackage{color}

\def\simge{\lower0.7ex\hbox{$\ \overset{>}{\sim}\ $}}
\def\simle{\lower0.7ex\hbox{$\ \overset{<}{\sim}\ $}}
\newcommand{\dg}{\dagger}

\begin{document}

\title{Quantum impurity in a Tomonaga-Luttinger liquid: continuous-time quantum Monte Carlo approach
}

\author{K. Hattori$^{1,2}$}
\email{hattori@issp.u-tokyo.ac.jp}
\author{A. Rosch$^1$}%
\affiliation{%
$^1$Institut f\"{u}r Theoretische Physik, Universit\"{a}t zu K\"{o}ln,
Z\"{u}lpicher Str., 77, D-50937 K\"{o}ln, Germany\\
$^2$Institute for Solid State Physics, University of Tokyo, 5-1-5, Kashiwanoha,\ Kashiwa,\ Chiba 277-8581, Japan
}%

\date{\today}

\begin{abstract}
We develop a continuous-time quantum Monte Carlo (CTQMC) method for
 quantum impurities coupled to interacting quantum wires described by a 
 Tomonaga-Luttinger liquid. The method is
 negative-sign free for any values of the Tomonaga-Luttinger parameter,
 which is rigorously proved, 
and thus, 
 efficient low-temperature calculations are possible. 
Duality between electrons and bosons in one dimensional systems allows
 us to construct a simple formula for the CTQMC algorithm in these systems.
 We show that the CTQMC for Tomonaga-Luttinger liquids can be implemented with only minor modifications of previous CTQMC codes developed  for impurities coupled to non-interacting fermions.
We apply this
 method to the Kane-Fisher model of a potential scatterer in a spin-less
 quantum wire and 
to a single spin coupled with the edge state of a two-dimensional topological insulator assuming an anisotropic XXZ coupling.
 Various dynamical response functions such as the
 electron 
 Green's function and spin-spin correlation functions are calculated numerically
 and their scaling properties are discussed.
\end{abstract}
\maketitle

\section{Introduction}
Electronic correlations
play fundamental roles in determining low-energy phenomena in
one-dimensional electron systems \cite{TLL,Giam}. 
Bosonization is a powerful technique to treat such correlations
exactly. 
In the presence of impurities, however, 
it is well known that impurities in one-dimensional systems drastically influence transport
properties of the systems.
Such an
example has been found in a classical model by Kane and Fisher in their 
pioneering work about a backward scattering potential problem in a spinless
quantum wire \cite{KF}. There, for the case of repulsive interaction 
 [Tomonaga-Luttinger (TL) parameter $g<1$], the conductance $G$ vanishes at
 zero temperature ($T=0$) and the potential barrier becomes infinitely strong
 and cut the wire into two parts, while for attractive cases with $g>1$,
 the potential becomes zero in the low-energy limit and 
 there remains a finite value of the conductance at $T=0$. 

For acquiring knowledge about thermodynamic, transport, and dynamical properties of such
systems, bosonization combined with perturbative renormalization group
methods 
\cite{KF,Delft,Furusaki}, Bethe
ansatz \cite{Bethe}, and 
{functional renormalization group \cite{fRG}} have been intensively used so far. 
To obtain numerically exact results for bosonized impurity problems, 
path integral Monte Carlo approaches have been employed
\cite{PIMC1,PIMC2,PIMC3}. A bosonic numerical renormalization group
method \cite{nrg} is also a powerful technique to investigate their
low-energy properties. 
While they are very useful approaches, there is still need for even 
more powerful numerical approaches that allow to compute wide range of 
temperature properties and dynamical correlation functions even for more complex models in an exact way.

To this end, in this paper, we will develop a continuous-time quantum
Monte Carlo (CTQMC)
method \cite{Rubtsov, Werner, Otsuki, Gull} in the Tomonaga-Luttinger liquid (TLL) in one-dimensional systems
coupled to an impurity. The CTQMC previously has been developed to describe quantum
impurities 
coupled to {\em non-interacting} environments. 
It has mainly been used for fermionic systems 
 and extensively used in the frame work of the
dynamical mean field theory \cite{DMFT} as an exact numerical solver for the
effective-impurity problem in it. Recent development \cite{Anders,Otsuki-b} of the algorithm
also enables us to treat bosonic systems and mixture of bosons and fermions. The advantage of the CTQMC is
that this allows us to calculate various quantities at low temperatures
in efficient ways and for some simple models there is no negative sign
problem \cite{Werner,Otsuki}. 

Our algorithm of CTQMC for TLL has advantages in the following points.
(i) Bosonization allows us to treat correlation arising from strong
interactions in the environment exactly. 
(ii) There is no negative sign problem for any parameters, which is,
indeed, proved analytically. This enables us to carry out
low-temperature analysis with high precision.
(iii) There are close relations to the
fermionic version of CTQMC, although the whole algorithm is written in
the bosonization language. This enables ones to implement the CTQMC for
the TLL easily from their fermionic CTQMC code.
(iv) The method can be applicable to not only potential scattering
problems but also to Kondo-type problems \cite{Wu,Mac0,Mac} without a negative sign problem.
(v) The electron Green's functions, the boson-boson correlations, conductance, the spin-spin
correlation functions, and various local correlators are calculable. 
 (vi) Compared with lattice QMCs, our formalism is free from finite size
effects at low temperatures and phase spaces for the random walk is expected
to be much smaller, and, thus, the computational cost is much lower.

This paper is organized as follows.
In Sec. \ref{model}, we will explain the models used in this paper. Section
\ref{secCTQMC} will be devoted to illustrate our algorithm of the CTQMC
for TLL. The method will be applied to the Kane-Fisher model \cite{KF} in
 Sec. \ref{KFmodel} and the XXZ Kondo problem \cite{Wu,Mac0,Mac} in Sec. \ref{XXZ}. We will
discuss possible extension in Sec. \ref{dis} and summarize the present
results in Sec. \ref{sum}.

\section{Models}\label{model}
In this section, we will introduce our model. First, we will show a 
one-dimensional Tomonaga-Luttinger liquid Hamiltonian and explain our
notation of the bosonization we will use throughout this paper.
The second part is an introduction of impurity-electron interactions. We
will use a general expression that can be used in two models we will
discuss in Sec. \ref{appli}.
\subsection{One-dimensional bulk Hamiltonian}

We consider spin-less fermions in one-dimensional (1D) systems
 whose
non-interacting Hamiltonian is given by \cite{Giam}
\begin{eqnarray}
H_{1D}&=&\frac{iv_F}{2\pi}\!\!\int_{-\frac{l}{2}}^{\frac{l}{2}}\!\!dx
\!:\!\Big\{ \psi^{\dagger}_{L}(x)\partial_x
 \psi_{L}(x)
-
\psi^{\dagger}_{R}(x)\partial_x
 \psi_{R}(x)
\Big\}\!:,\ \ \ \
\label{H0}
\end{eqnarray}
 where $\psi^{\dag}_{L,R}(x)$ is the fermion creation operator at the
 position $x$ and $L$($R$) refer to
left(right)-moving component. :$A$: indicates the normal ordering of the
operator $A$, 
$l$ is the system size, and $v_F$ is the Fermi velocity.
The fermion field $\psi_{L,R}(x)$ satisfies the anticommutation relation
\begin{eqnarray}
\{\psi_{\rho}(x),\psi^{\dagger}_{\rho'}(x')\}=2\pi\delta_{\rho\rho'}\delta(x-x'). \label{antiPsi}
\end{eqnarray}

Following the standard bosonization procedure \cite{Giam,Delft}, we define bosons
$\phi_{L,R}(x)$ as
\begin{eqnarray}
\psi_{L,R}(x)=a^{-1/2}F_{L,R}e^{i\phi_{L,R}(x)}, \label{bosformula}
\end{eqnarray}
where $a$ is the short-distance cutoff. 
The Bose fields satisfy
\begin{eqnarray}
[\phi_{\rho}(x),\partial_{x'}\phi_{\rho'}(x')]=2\pi i \delta_{\rho\rho'}\Big[\delta(x-x')-\frac{1}{2l}\Big],
\end{eqnarray}
where the $O(l^{-1})$ term is explicitly written.
Two Klein factors $F_L$ and $F_R$ have been 
introduced in Eq. (\ref{bosformula}) to reproduce the anticommutation relation of
$\psi_{R,L}$ [Eq.~(\ref{antiPsi})]. Their anticommutation relation is 
\begin{eqnarray}
\{F_{\rho},F^{\dagger}_{\rho'}\}=2\delta_{\rho\rho'} \quad {\rm with}\ \
F^{\dag}_{\rho}F_{\rho}=F_{\rho}F^{\dag}_{\rho}=1,
\end{eqnarray}
and 
\begin{eqnarray}
\{F_{\rho}^{\dagger},F^{\dagger}_{\rho'}\}=\{F_{\rho},F_{\rho'}\}=0,\ \ 
 {\rm for}\ \rho\ne \rho'.
\end{eqnarray}
Note that $F_{\rho}F_{\rho}\ne 1$, and the two bosons are independent
fields commuting with each other and with the Klein factors, which is a physically correct description as is evident from
the definition of $L$ and $R$ \cite{Delft}.

Electron-electron interactions are easily taken into account in the
bosonized theory and then the bosonized
Hamiltonian reads
\begin{eqnarray}
H_{1D}=\frac{v}{4}\int_{-\frac{l}{2}}^{\frac{l}{2}} \frac{dx}{2\pi}
 :\Big\{\frac{1}{g}\Big[\partial_x
 \phi_-(x)\Big]^2+g\Big[\partial_x\phi_+(x)\Big]^2\Big\}:,\ \ \ 
\label{H0b}
\end{eqnarray}
with $\phi_{\pm}(x)=\phi_L(x)\pm\phi_R(x)$ and $g$ is the
TL parameter that characterizes the bosonic theory:
$g=1$ corresponds to the noninteracting case and
$0<g<1$ ($g>1$) describes repulsive (attractive) interactions, respectively. The
velocity $v_F$ is now renormalized as $v\equiv v_F/g$. 
Throughout this paper, we will be interested in the repulsive case.

For later purposes, we introduce another representation following Delft
and Schoeller as \cite{Delft}
\begin{eqnarray}
\Phi_{\pm}(x)&=&
\frac{1}{2\sqrt{2}}\Bigg\{
\Big(\frac{1}{\sqrt{g}}+\sqrt{g}\Big)
\Big[\phi_L(x)\mp\phi_R(-x)\Big]\nonumber\\
&&\quad\quad\pm
\Big(\frac{1}{\sqrt{g}}-\sqrt{g}\Big)
\Big[\phi_L(-x)\mp\phi_R(x)\Big]
\Bigg\}.
\end{eqnarray}
Then, the Hamiltonian (\ref{H0b}) is rewritten as 
\begin{eqnarray}
H_{1D}=\frac{v}{2}\int_{-\frac{l}{2}}^{\frac{l}{2}} \frac{dx}{2\pi}
 :\Big\{\Big[\partial_x
 \Phi_-(x)\Big]^2+\Big[\partial_x\Phi_+(x)\Big]^2\Big\}:.\label{H0_2}
\end{eqnarray}
At $x=0$, a simple relation holds \cite{Delft}, 
\begin{equation}
\Phi_{\pm}\equiv\Phi_{\pm}(0)=\frac{g^{\mp 1/2}}{\sqrt{2}}\big[\phi_L(0)\mp\phi_R(0)\big].
\end{equation}

\subsection{Impurity potentials}
Now, we introduce interactions between a quantum impurity located at $x=0$ and the
interacting electrons of the TLL. We consider a coupling by
single-particle scattering (generalization to more complicated
interactions is straightforward  but, of course, model dependent).
The interactions, $V=V^{\sigma}_F+V_B$, are decomposed into two parts,  a 
forward-scattering channel, $V^{\sigma}_F$, and a backward-scattering channel described by $V_B$,
\begin{eqnarray}
V^{\sigma}_F&=&\lambda_F :\Big[
 \psi^{\dg}_L(0)\psi_L(0)-\sigma\psi^{\dg}_R(0)\psi_R(0)\Big]:{\hat{X}}^{\sigma}_F,\label{HimpF}\\
V_B&=&\lambda_{B} \psi^{\dg}_L(0)\psi_R(0)\hat{X}_{B}
+{\rm H.c.}, \label{Himp}
\end{eqnarray}
The scattering of electrons can change the state of the impurity (e.g.,
flip a spin). This is described by the impurity operators
$\hat{X}^{\sigma=\pm}_{F}$ and $\hat{X}_B$. They will be discussed in
later sections. In the most general case, their form can be derived
by first bosonizing the model and then identifying the two terms
discussed above by comparing them to the bosonized version of
Eqs. (\ref{HimpF}) and (\ref{Himp}) discussed below.
Note that $\lambda_{F,B}$ has the
dimension of [energy]$\times$[length] and $\hat{X}$'s are dimensionless operators.

In terms of the bosons, Eqs. (\ref{HimpF}) and (\ref{Himp}) read
\begin{eqnarray}
V^{\sigma}_F&=&\lambda_F \sqrt{\frac{2}{g^{\sigma}}}
 \partial_x\Phi_{\sigma}(0)\hat{X}^{\sigma}_F,\\
V_B&=&a^{g-1}\lambda_B F^{\dg}_LF_R
 \Big(a^{-g}e^{i\sqrt{2g}\Phi_+}\Big)\hat{X}_B\nonumber\\
&&+a^{g-1}\lambda^*_B F^{\dg}_RF_L
 \Big(a^{-g}e^{-i\sqrt{2g}\Phi_+}\Big)\hat{X}^{\dg}_B\\
&\equiv&\tilde{\lambda}_B F_L^{\dg}F_R \hat{V}_{+\sqrt{2g}}(\Phi_+)\hat{X}_B^+
+\tilde{\lambda}^*_B F_R^{\dg}F_L \hat{V}_{-\sqrt{2g}}(\Phi_+)\hat{X}_B^-,\nonumber\\
\label{HimpB}
\end{eqnarray}
where $\hat{X}^+_B=\hat{X}_B$, $\hat{X}_B^-=\hat{X}_B^{\dg}$, and $\tilde{\lambda}_B=a^{g-1}\lambda_B$. The vertex operator is
defined as 
\begin{eqnarray}
V_{\pm \sqrt{2g}}(\Phi_+)=a^{-g}\exp \Big( \pm i\sqrt{2g}\Phi_+\Big).
\end{eqnarray}
This normalization of the vertex operator leads to following the bare  (i.e., in the absence of the impurity) 
two-point
correlator as a function of imaginary time $\tau$, 
\begin{eqnarray}
\langle V_{\sqrt{2g}}(\Phi_+,\tau)V_{-\sqrt{2g}}(\Phi_+,\tau')\rangle =|\tau-\tau'|^{-2g},\label{twopoint}
\end{eqnarray}
 at $T\to 0$ for $a\to 0$ and $l\to \infty$ [see also the definition of
multipoint correlators in Eq. (\ref{VVV}) below]. 
\section{Continuous-time Quantum Monte Carlo Method}\label{secCTQMC}
In this section, we will explain how continuous-time quantum Monte Carlo
method can be applied to the impurity problem in the TLLs. 
We will demonstrate that the configuration weight for a given
snapshot is easily calculated by the technique developed in fermionic
CTQMCs. We will therefore omit detailed explanations about update operations,
since these are essentially the same as in the fermionic CTQMCs \cite{Gull}.

\subsection{Partition function}
We want to evaluate the partition function $Z$,
\begin{eqnarray}
Z={\rm Tr} \exp[-\beta (H_0+V)],
\end{eqnarray}
within a Monte
Carlo approach.
Here, the ``noninteracting'' part $H_0$ is the sum of the
one-dimensional TLL 
and the local impurity Hamiltonian, $H_0=H_{1D}+H_{\rm imp}$. In this paper, we
will analyze models with $H_{\rm imp}=0$ (e.g., a magnetic impurity in the absence of magnetic fields). 
Via perturbative expansion of $V$, we can express $Z$ as
\begin{eqnarray}
\frac{Z}{Z_0}=\Bigg\langle T_{\tau} \exp\Bigg[-\int_0^{\beta} V(\tau)d\tau\Bigg]\Bigg\rangle_0,
\end{eqnarray}
where $Z_0={\rm Tr} e^{-\beta H_0}$ and $\langle A \rangle_0=[{\rm Tr}
 A e^{-\beta H_0}]/Z_0$ and $T_{\tau}$ indicates the time-ordered
 product. In this paper, we will discuss situations where the forward-scattering part ($\lambda_F$) can
be eliminated by an appropriate unitary transformation or where $\lambda_F=0$ due
 to symmetry requirements. Thus, we retain only $V_B$ and 
in order to distinguish the two terms in $V_B$, we define
\begin{eqnarray}
v_B^+&\equiv&\tilde{\lambda}_B F_L^{\dg}F_R \hat{V}_{+\sqrt{2g}}(\Phi_+)\hat{X}_B^+,\\
v_B^-&\equiv&\tilde{\lambda}^*_B F_R^{\dg}F_L \hat{V}_{-\sqrt{2g}}(\Phi_+)\hat{X}_B^{-}.
\end{eqnarray}
A general $N$th order term $\delta Z_N$ in the partition function is
expressed as
\begin{eqnarray}
\delta Z_N &=& \frac{(-1)^N}{N!}\int_0^{\beta} d\tau_1 \cdots \int_0^{\beta} d\tau_N\nonumber\\
&&\quad\quad\times\langle T_{\tau}V_B(\tau_1)V_B(\tau_2)\cdots V_B(\tau_N) \rangle_0. \label{dZN}
\end{eqnarray}
Due to the fermion number conservation encoded by the Klein factors, the number of $v_B^+$ in
Eq. (\ref{dZN}) has to be the same as that for $v_B^-$, and therefore 
only even $N=2k$ terms with $k \in \mathbbm Z$, contribute.
Now, consider a fixed series of times $\{\tau;\tau_1>\tau_2>\cdots>\tau_N\}$.
Then $\delta Z_N\{\tau\}$ becomes 
\begin{eqnarray}
\delta Z_N \{\tau\}&=& 
|\tilde{\lambda}_B|^{2k}\langle v^{\sigma_1}_B(\tau_1)
v_B^{\sigma_2}(\tau_2)\cdots v_B^{\sigma_{2k}}(\tau_{2k})\rangle_0, \ \
\ \label{dZN-1}
\end{eqnarray}
with $\sigma_1$, $\sigma_2$, $\cdots, \sigma_{2k}=+$ or $-$. The
 partition sum is obtained by averaging over all $\sigma_i$, all times,
 and all $k$ using a Monte Carlo procedure.

Since the product of Klein factors gives factor unity, we can write
Eq. (\ref{dZN-1}) as a product of  a TLL correlator and a correlator involving only impurity operators
\begin{eqnarray}
\delta Z_{2k}\{\tau\} &=& |\tilde{\lambda}_B|^{2k} \delta Z_{2k}^{\Phi_+}\{\tau\}\delta Z_{2k}^{X}\{\tau\}.\label{dZ}
\end{eqnarray}
In the following subsections, we will analyze the two sectors in details. 
\subsection{Impurity average}
Here, we discuss the local part $\delta Z_{2k}^X$. 
$\delta Z_{2k}^X$ is the time-ordered product 
of $\hat{X}_B^{\pm}$'s:
\begin{eqnarray}
\delta Z_{2k}^{X}\{\tau\}=\langle \hat{X}_B^{\sigma_1}(\tau_1) \cdots
 \hat{X}_B^{\sigma_{2k}}(\tau_{2k})\rangle_{\rm imp}.
\end{eqnarray}
Here, $\langle \cdot \rangle_{\rm imp}$ is the average with respect to
the impurity Hamiltonian. 
As noted above, both $\hat{X}^+_B$ and $\hat{X}_B^-$ appear $k$ times
and, for later convenience, we define new $\tau$ indices $\tau_i^{\pm}$
with $1\le i \le k$, such that the operators $\hat{X}^+_B$ ($\hat{X}_B^-$)  are evaluated at the times $\tau_i^+$ ($\tau_i^-$) with  time-ordering in each index, $\tau_i^{\pm}>\tau_{i+1}^{\pm}$.
\subsection{Boson average}
For the bosonic part, $\delta Z_{2k}^{\Phi_+}$ is the time-ordered $2k$-point
correlation function of $\hat{V}_{\pm \sqrt{2g}}(\Phi_+)$:
\begin{eqnarray}
\delta Z_{2k}^{\Phi_+}\{\tau\}\!=\!\langle
 \hat{V}_{\sigma_1 \sqrt{2g}}(\Phi_+,\tau_1^{\sigma_1}) \cdots
 \hat{V}_{\sigma_{2k} \sqrt{2g}}(\Phi_+,\tau_{2k}^{\sigma_{2k}})
 \rangle_{\Phi_+}.\ \ \ \ \
\end{eqnarray}
Here, the  boson average $\langle \cdot \rangle_{\Phi_+}$ is evaluated using the
Gaussian TLL Hamiltonian (\ref{H0_2}). 
It is well known that the correlation function of vertex operators,  
\begin{eqnarray}
\hat{V}_{\lambda_i}(\Phi_+,\tau_i)=a^{- \lambda_i^2/2} e^{i\lambda_i
 \Phi_+(\tau_i)}, 
\end{eqnarray}
 are
calculated as \cite{Delft}
\begin{eqnarray}
&&\langle T_{\tau}
\hat{V}_{\lambda_1}(\Phi_+,\tau_1) \cdots
\hat{V}_{\lambda_N}(\Phi_+,\tau_N)\rangle_{\Phi_+}\nonumber\\
&&=
\Big(\frac{2\pi}{l}\Big)^{\frac{1}{2}\big(\sum_j^N
\lambda_j\big)^2}\prod_{i<j}^N \Big[s(\tau_{ij})\Big]^{\lambda_i\lambda_j},
\label{VVV}
\end{eqnarray}
with
\begin{eqnarray}
s(\tau_{ij})\equiv \frac{v\beta}{\pi}\sin \Big[
\frac{\pi}{v\beta}\big(v|\tau_{ij}| +\epsilon(|\tau_{ij}|) \big)
\Big].\label{VVVV2}
\end{eqnarray}
Here, $\tau_{ij}=\tau_i-\tau_j$, and, in order to prevent the divergence,
the cutoff function $\epsilon(\tau)$
is necessary and it satisfies 
\begin{eqnarray}
   \epsilon(\tau)=-\epsilon(\beta-\tau),\quad \epsilon(0)=a, \quad {\rm
    and}\ \ \epsilon(\beta)=-a.
\end{eqnarray}
 In actual calculations, we will use the following function
$\epsilon(\tau)$ throughout this paper: 
\begin{eqnarray}
\epsilon(\tau)=a \ \!{\rm sgn}(\beta/2-\tau).
\end{eqnarray}
For very high temperatures (not considered in this paper), it is sometimes useful to use a smooth function
in order to remove the discontinuity appearing in physical quantities 
such as
\begin{eqnarray}
\epsilon(\tau)=a \tanh \Bigg[c \frac{\beta/2-\tau}{\tau(\beta-\tau)}\Bigg],
\end{eqnarray}
with $c$ being a positive constant. 
In the $l\to \infty$ limit, Eq. (\ref{VVV}) vanishes unless 
$\sum_j\lambda_j=0$. Thus, a ``neutrality condition,'' $\sum_j\lambda_j=0$, 
has to be fulfilled. In our case, this is automatically enforced by the fermion
number conservation and we obtain 
\begin{eqnarray}
\delta Z_{2k}^{\Phi_+}\{\tau\} = 
\prod_{i<j}^{2k} \big[s(\tau_{ij})\big]^{\lambda_i\lambda_j}>0,\label{VVV-Llimit}
\end{eqnarray}
with $\lambda_{i,j}=\pm \sqrt{2g}$. An important observation is that  Eq. (\ref{VVV-Llimit}) is
positive definite, and, thus, our Monte Carlo method is negative-sign free
if $\delta Z_{2k}^X>0$. 

Equation (\ref{VVV-Llimit}) can be further simplified via the
``generalized'' Wick's theorem \cite{Delft}, which is valid if and only if $a=0$.
We utilize this theorem, although actual numerical calculations are done
with finite $a$. The theorem might be most easily obtained by comparing
the partition function for noninteracting spinless fermion in one
dimension and that in the bosonization representation. The result is 
\begin{eqnarray}
\delta Z_{2k}^{\Phi_+}\{\tau\} = 
|{\rm det} \hat{S}_k\{\tau\}|^{2g}.\label{VVV-Llimit2}
\end{eqnarray}
The $k\times k$ matrix $\hat{S}_k$ is given by
\begin{eqnarray}
\Big[\hat{S}_k\{\tau\}\Big]_{ij}=-{\rm
 sgn}(\tau_{ij})\big[s(\tau_{ij})\big]^{-1}, \quad 1\le i,j \le k, \ \ 
\end{eqnarray}
and the index $i(j)$ corresponds to $\tau_i^-(\tau_j^+)$. This form is
particularly useful, since we can use the fast-update algorithm developed
in the conventional fermionic CTQMC methods \cite{Rubtsov}.

\section{Applications}\label{appli}
In this section, we will apply our CTQMC method to two models. One is 
the Kane-Fisher model describing a backward-scattering impurity potential in a (spinless)
quantum wire \cite{KF}. The other is the XXZ Kondo problem \cite{Wu,Mac0,Mac} in
helical liquids, i.e., on the edge of two-dimensional topological insulators.
\subsection{Kane-Fisher model}\label{KFmodel}
The Kane-Fisher model is defined by considering forward scattering
$\hat{X}^{\sigma}_F$ with $\sigma=-$ and $\hat{X}_F^-=1$ in
Eq. (\ref{HimpF}) and by setting  $\hat{X}_B=\hat{X}_B^{\dg}=1$ in Eq. (\ref{Himp}) as a potential scatterer has no  internal degrees of freedom. Since $V_F$
contains only $\Phi_-$ and $V_B$ only $\Phi_+$, we can separately
analyze the two. The $V_F$ part is  trivial because it can be
absorbed into $\Phi_-$ terms in $H_{1D}$ by a unitary
transformation \cite{Delft}. Thus, in the following, we analyze $V_B$ part in
details. Note that because $\hat{X}_B=1$ and thus $\delta Z_{2k}^X=1$,
we can use the positivity of (\ref{VVV-Llimit})  to conclude immediately that there is no negative-sign problem. Throughout this subsection, we
set $v=v_F/g$ and fix $v_F/\xi=1$ for the unit of energy, where $\xi=1$ is
the relevant microscopic unit of length, which is, for example, set by the typical width of the potential.

 The model itself has been extensively analyzed by various
authors \cite{KF,Delft,Furusaki,Bethe,fRG,PIMC1,PIMC2}, and now its
low-energy properties are well understood. We will
study this problem as a benchmark of our algorithm. We will discuss a physical
quantity that has not been investigated so far: the electron Green's
function in imaginary time. This is the most natural quantity for
imaginary-time algorithms like CTQMC. Though it is not a directly measurable quantity in experiments, the
numerically-exact results can be used to obtain the density of states by
using analytical continuation techniques \cite{Jarrel} (not covered in this paper).

\subsubsection{Electron Green's function}
Let us consider the  local Green's function for $\psi_L(\tau>0,x=0)$,
\begin{eqnarray}
G_L(\tau)&=&-\langle \psi_L(\tau) \psi_L^{\dagger}(0)\rangle\\
\!\!\!\!&=& -a^{\frac{g}{2}+\frac{1}{2g}-1}
\langle \hat{V}_{-\frac{1}{\sqrt{2g}}}(\Phi_-,\tau) \hat{V}_{\frac{1}{\sqrt{2g}}}(\Phi_-,0)\rangle_{\Phi_-}\nonumber\\
\!\!\!\!&&\times\langle F_L(\tau) F^{\dg}_L(0)\rangle_F \langle\hat{V}_{-\sqrt{\frac{g}{2}}}(\Phi_+,\tau)
\hat{V}_{\sqrt{\frac{g}{2}}}(\Phi_+,0)\rangle_{\Phi_+}\nonumber\\
\!\!\!\!&\equiv& -a^{\frac{g}{2}+\frac{1}{2g}-1}
G^-_L(\tau)G^+_L(\tau), \label{defGL+}
\end{eqnarray}
with
\begin{eqnarray}
G^-_L(\tau)&=&\langle \hat{V}_{-\frac{1}{\sqrt{2g}}}(\Phi_-,\tau)
 \hat{V}_{\frac{1}{\sqrt{2g}}}(\Phi_-,0)\rangle_{\Phi_-},\\
G^+_L(\tau)&=&\langle F_L(\tau)F_L^{\dagger}(0)\rangle_F \langle \hat{V}_{-\sqrt{\frac{g}{2}}}(\Phi_+,\tau)
 \hat{V}_{\sqrt{\frac{g}{2}}}(\Phi_+,0)\rangle_{\Phi_+}.\nonumber\\
\end{eqnarray}
Here, $\langle \cdots \rangle_F$ is the average over Klein factors.
The correlation function for 
the $\Phi_-$ part $G^-_L(\tau)$ is trivial, leading to 
\begin{equation}
G^-_L(\tau)= \big[s(\tau)\big]^{-\frac{1}{2g}}.
\end{equation}
Here $s(\tau)$ is given by Eq. (\ref{VVVV2}). 
 Note that for our model nonlocal Green's function can be
expressed in terms of local correlators as the bosons of the TLL are noninteracting.
In the following, we explain how one can calculate the $\Phi_+$ and the
Klein factor parts $G_L^+(\tau)$ in our CTQMC. 

In the Monte Carlo simulations, time-ordered averages of an operator
$\hat{A}$ is estimated as
\begin{eqnarray} 
\langle T_{\tau}\hat{A}\rangle=\frac{1}{N_{\rm MC}}\sum_{m=1}^{N_{\rm
 MC}} \frac{\langle T_{\tau}
 \hat{A}\delta\hat{Z}_{N_m}\{\tau\} \rangle_0}{\delta Z_{N_m}\{\tau\} }, \label{Av}
 \end{eqnarray}
where $N_{\rm MC}$ is the number of Monte Carlo samplings and 
we have defined an operator form of $\delta Z_{N_m}\{\tau\}$, see 
Eq.~(\ref{dZN-1}). For $N_m=2k$, this is given by
\begin{eqnarray}
\delta\hat{Z}_{2k}\{\tau\} = |\tilde{\lambda}_B|^{2k}
v_B^{\sigma_1}(\tau_1)v_B^{\sigma_2}(\tau_2)\cdots v_B^{\sigma_{2k}}(\tau_{2k}).
\end{eqnarray}

For the Green's function $G^+_L(\tau_{ij})$, we need to calculate
Eq. (\ref{Av}) with
$
\hat{A}=F_L(\tau_i)\hat{V}_{-\eta}(\Phi_+,\tau_i) F_L^{\dg}(\tau_j) \hat{V}_{\eta}(\Phi_+,\tau_j)
$ with $\eta=\sqrt{g/2}$. 
As derived in Appendix \ref{deriG}, we need to sample the following quantity for $\tau_{ij}>0$,
\begin{eqnarray}
 {\mathcal G}_{i>j}^{(2k)}&=&
(-1)^{P_{ij}}\big[s(\tau_{ij})\big]^{\frac{g}{2}}\Bigg|
\frac{{\rm det} \hat{S}_{k+1}\{\tau\oplus\tau_i,\tau_j\}}{{\rm det} \hat{S}_{k}\{\tau\}}
\Bigg|^g.\label{Gfin}
\end{eqnarray}
Here, $P_{ij}$ is the number of vertices between $\tau_i$ and $\tau_j$
in the MC snapshot and a similar expression is obtained for
$\tau_{ij}<0$. The notation $\{\tau \oplus \tau_i,\tau_j\}$ represents 
that $\tau_i$ and $\tau_j$ are added to $\{\tau\}$.
Note that to derive Eq. (\ref{Gfin}), we have used the generalized
Wick's theorem mentioned before.  
Then we obtain
\begin{eqnarray}
G^+_{L}(\tau_{ij})&=&
\big\langle
 {\mathcal G}_{i>j}^{(2k)}
\big\rangle, \quad {\rm for}\ \tau_{ij}>0,  \label{GLplus1}
\end{eqnarray}
and a similar expression is applied to $G^+_L(\tau_{ij}<0)=-G^{+}_L(\beta+\tau_{ij})$.

An alternative approach is, however, more efficient in regimes where
high orders of perturbation theory are needed. 
Following Refs.~\cite{Rubtsov,Werner},   we can also derive an alternative expression for calculating the Green's
function $G^+_L(\tau)$. The quantity that corresponds to
Eq. (\ref{Gfin}) is now given by
\begin{eqnarray}
 \tilde{\mathcal G}_{i>j}^{(2k)}
&=&\frac{(-1)^{P_{ij}}}{|\tilde{\lambda}_B|^{2}}\big[s(\tau_{ij})\big]^{\frac{g}{2}}
\Bigg|\frac{
{\rm det} \hat{S}_{k-1}\{\tau \ominus \tau_i^-,\tau_j^+\}
}{
{\rm det} \hat{S}_{k}\{\tau\}
}
\Bigg|^{g}.\ \ \ \ \ \label{G2}
\end{eqnarray}
Note that in Eq. (\ref{G2}), $\tau_i^-$ and $\tau_j^+$ are chosen in a
given snapshot $\{\tau\}$, while in Eq. (\ref{Gfin}), $\tau_{i}$ and
$\tau_j$ are external ones. This implies that by computing one snapshot with $k$ pairs of time variables
one obtains contribution to the Green's function for about $k^2$ different $\tau_{ij}$, which helps to reduce the statistical error. The notation $\{\tau \ominus
\tau_i,\tau_j\}$ represents that $\tau_i$ and $\tau_j$ are removed from $\{\tau\}$.

Since the ratio of two determinants in Eq. (\ref{G2}) is simply
$(\hat{S}_k^{-1}\{\tau\})_{ji}$, which is calculated in every MC
process, this also reduces computational costs \cite{Rubtsov,Werner}. We can also derive a similar
expression for $\tau_i<\tau_j$, $\tilde{
\mathcal G}_{i<j}^{(2k)}$.
Summing over all possible combinations $(i,j)$ for a given snapshot at
$2k$th order and dividing by $\beta$, we obtain
\begin{eqnarray}
G^+_{L}(\tau)\!&=&\!\frac{1}{\beta}
\Big\langle
\sum_{ij}^k\Big[
 \tilde{\mathcal G}_{i>j}^{(2k)}\delta(\tau_{ij}-\tau)
-\tilde{\mathcal G}_{i<j}^{(2k)}\delta(\beta+\tau_{ij}-\tau)
\Big]
\Big\rangle. \nonumber\\
\label{GLplus2}
\end{eqnarray}
The two alternative formulas (\ref{GLplus1}) and (\ref{GLplus2}) can be used for
checking the program code.

\subsubsection{Bench mark for $g=1$}\label{a-dep}
\begin{figure}[t!]
\begin{center}
    \includegraphics[width=0.4\textwidth]{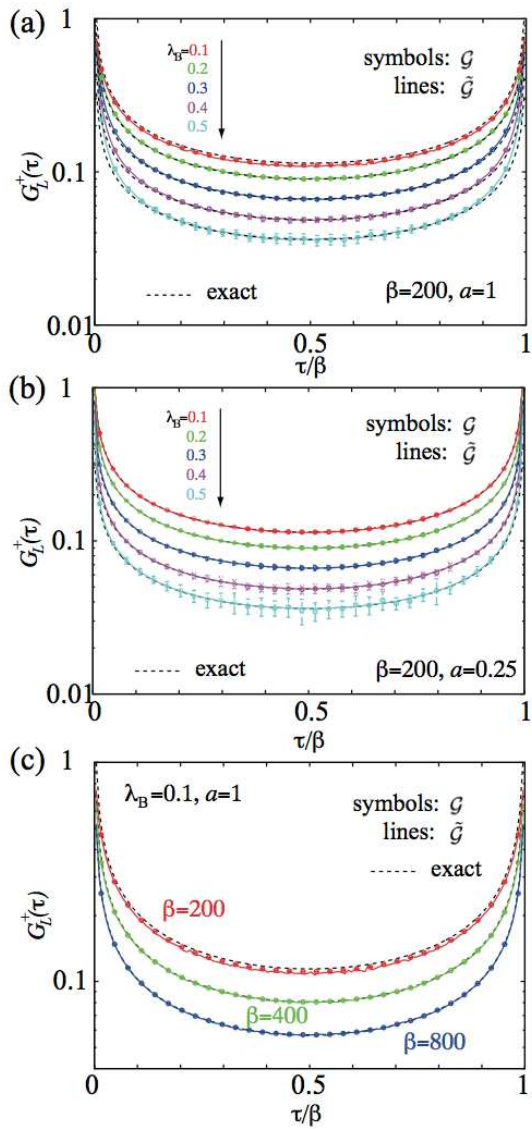}
\end{center}
\caption{(Color online) $G_L^+(\tau)$ vs $\tau/\beta$ for several coupling constants
 $\lambda_B=0.1$(top)-$0.5$(bottom) and inverse temperatures $\beta$.
The exact result (dashed line) is compared to  two numerical methods [points with error bars: Eq. (\ref{Gfin}), solid lines: Eq. (\ref{G2})]. A comparison 
of panel (a) (cutoff $a=1$) to panel (b) (cutoff $a=0.25$) shows that small deviations from the exact result vanish for small $a$. Panel (c) shows that highly accurate results can be obtained even for very low $T$.}
\label{fig-benchmark}
\end{figure}

In this subsection, we show the results for $g=1$, i.e.,
a system of noninteracting electrons. We compare the electron Green's
function obtained in the CTQMC and the exact results as follows:
\begin{eqnarray}
G^{+, \rm ex}_L(\tau)=\frac{\big[s(\tau)\big]^{-\frac{1}{2}}}{1+ \pi^2\lambda_B^2/v^2},\label{exact}
\end{eqnarray}
which can be easily obtained from the equations of motion for the Green's
functions.

Figure \ref{fig-benchmark} shows $G^+_L(\tau)$ as a function of $\tau$ for $\beta=200$
and several
parameters $\lambda$ and $a$. In each plot, the exact result (dashed lines) and the result of the two methods described above are shown. The points with error bars (indicating the statistical error arising from the Monte Carlo sampling) are obtained from 
Eq. (\ref{Gfin}), while the solid lines have been calculated from  Eq. (\ref{G2}) (the statistical error can be read off from the size of the noise in the curves).  
As one can
see, the numerical data and the exact results are consistent with each
other. More precisely, the Green's functions are  only identical in the
limit $a \to 0$ [we used this limit both in the derivation of the CTQMC
approach and in Eq.~(\ref{exact})]. 
Figure \ref{fig-benchmark}  shows that tiny systematic deviations
of the exact and the numerical result visible for $a=1$ become smaller than the noise for $a=0.25$. In the following, we will always use $a=1$ as the universal properties 
for $T\ll v/a$ and $\tau \gg a/v$ discussed in the following are independent of the cutoff.

Figure \ref{fig-benchmark} shows that highly accurate results are also obtained for low $T$.
Comparing the two computational methods (using the same computational time), we first note that both give reliable results. Which method is preferable depends in general both on the perturbation order and the type of binning in time used to extract data. For the parameter regime used in our calculations, we found the second approach to be more efficient. For very high orders of perturbation theory and small number of bins, however, the first approach can 
beat the second one in efficiency. In Sec. \ref{disWC}, we will discuss that in regimes, where the nonlinearities are irrelevant (attractive interactions), the second method is inefficient.

\subsubsection{Universal scaling function for electron Green's function}

The main prediction of Kane and Fisher \cite{KF} is that for repulsive
interactions, $g<1$, 
even a weak impurity effectively cuts the chain: Electrons  scatter so
efficiently 
from the slowly decaying Friedel oscillations
that for $T \to 0$ and at the Fermi energy one obtains perfect
reflection.  
The fact that the impurity cuts
the quantum wire can also be measured by tunneling spectroscopy, i.e.,
by considering 
the local Green's function close to the impurity. Based on the
assumption that 
the wire is perfectly cut by the impurity,
one expects for $T=0$,  
\begin{equation}
G^+_L(\tau\to \infty )\sim \tau^{-1/(2g)}, \label{asmy}
\end{equation}
as has been derived by Furusaki \cite{Furusaki}. This prediction can be checked analytically for $g=1/2$, 
where an exact analytic result can be derived \cite{Delft}. Equation (\ref{asmy}) should be compared to $G^+_L(\tau\to \infty )\sim \tau^{-g/2}$ obtained for $\lambda_B=0$, in the absence of the impurity.

Note that for the computation of the physical electron Green's function one has to consider a further
contribution, $G_{LR}(\tau)=-\langle
T_{\tau}\psi_L(\tau)\psi_R^{\dagger}(0)\rangle$, in addition to $G_L(\tau)$. 
It is possible to  calculate $G_{LR}(\tau)$ using our approach, but we do not
discuss it here for simplicity.

From general scaling arguments and the analysis of Kane and Fisher \cite{KF}, one expects 
for a weak potential scatterer (small $\lambda_B$) a crossover from $G^+_L(\tau\to \infty )\sim \tau^{-g/2}$ to $G^+_L(\tau\to \infty )\sim \tau^{-1/(2g)}$ described by a universal (but $g$-dependent) scaling function $\mathcal F_g$, 
\begin{eqnarray}
G_L^+(\tau)\approx [s(\tau)]^{-g/2} \, {\mathcal F}_g(\,T^*\tau\,, T/T^*),\label{scaling}
\end{eqnarray}
 where $ [s(\tau)]^{-g/2}$ is the Green's function for $\lambda_B=0$, see Eq.~(\ref{VVVV}). All dependence on the strength $\lambda_B$ of the impurity potential is thereby encoded in the characteristic energy scale $T^*$ with 
\begin{equation}
T^*=\frac{v}{a} \left(\frac{\lambda_B}{v}\right)^{1/(1-g)}, \label{Tstar}
\end{equation}
for small $\lambda_B$. The universal scaling form (\ref{scaling}) is expected to be valid whenever $T^*$ is
much smaller than the cutoff energy $v/a$. For $T=0$ and weak $\lambda_B$, the short time dynamics is determined
by the noninteracting result, $  {\mathcal F}_g(x \to 0, 0)=1$, while  $  {\mathcal F}_g(x \to \infty, 0)\propto x^{(g-1/g)/2}$, see Eq.~(\ref{asmy}).

In the following, we show our CTQMC results, which confirm the expected
behavior and allow us 
to calculate the full scaling function describing the crossover from weak to strong coupling.
 To our knowledge, this is the first demonstration of the numerically-exact 
Green's function in this model.
 
\begin{figure}[t!]
\begin{center}
    \includegraphics[width=0.48\textwidth]{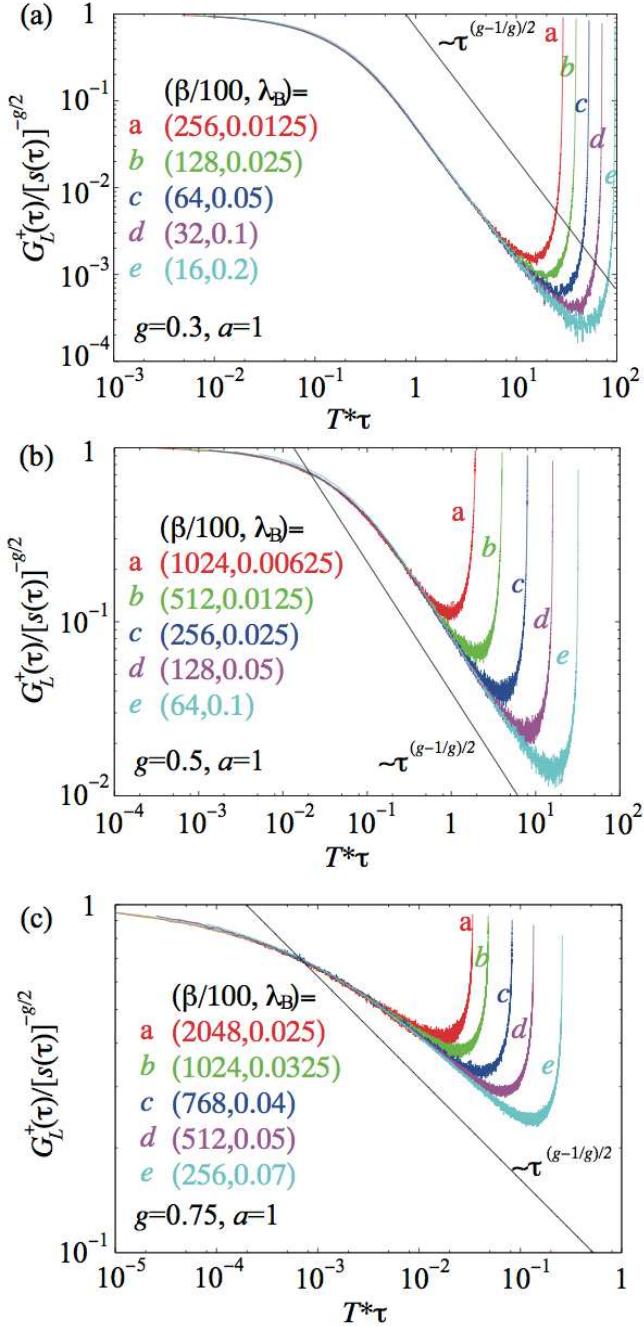}
\end{center}
\vspace{-0.5cm}
\caption{(Color online) $G_L^+(\tau)/[s(\tau)]^{-g/2}$ versus $T^*\tau$ for
 various parameter sets $(\beta,\lambda_B)$, $a=1$ and (a) $g=0.3$, (b)
 $g=0.5$, and (c) $g=0.75$. 
Straight lines show the $\tau^{(g-1/g)/2}$
 dependence expected from the fixed point where the impurity cuts the chain (the factor $\tau^{g/2}$ originates from the asymptotic form of $[s(\tau)]^{-g/2}$).}
\label{fig-Green_scale}
\end{figure}

Figure \ref{fig-Green_scale} shows $G^+_L(\tau)/[s(\tau)]^{-g/2}$ versus 
$T^*\tau$ for several parameter sets $(\beta,\lambda_B)$ and (a) $g=0.3$,
(b) $g=0.5$, and (c) $g=0.75$ for various temperatures $T$. 
Our numerical results 
reproduce the analytically expected behaviors: First, for wide ranges of $\lambda_B$ the curves  scale on top of each other (we have not used an appropriately rescaled temperature, therefore the upturns occur
at different points). Second, we obtain the analytically expected
asymptotic behavior with  $ {\mathcal F}_g(x \to 0, 0)=1$, while  $
{\mathcal F}_g(x \to \infty, 0)\propto x^{(g-1/g)/2}$, see 
Eq.~(\ref{asmy}). Third, our result provides the full crossover function
from weak to strong coupling.

\begin{figure}[t!]
\begin{center}
    \includegraphics[width=0.47\textwidth]{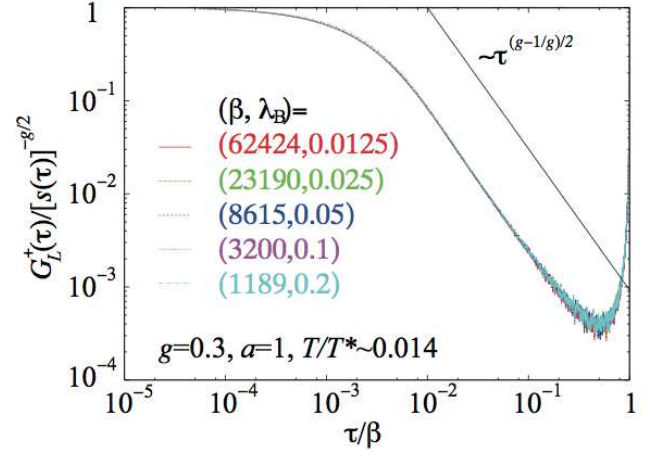}
\end{center}
\vspace{-0.5cm}
\caption{(Color online) $G_L^+(\tau)/[s(\tau)]^{-g/2}$ vs $\tau/\beta$ for $a=1$ and
 various parameters $(\beta,\lambda_B)$ keeping the ratio
 $T/T^*\simeq 0.014$ fixed.}
\label{fig-Green_T_T0}
\end{figure}

To prove that scaling works also at finite $T$, we show in
 Fig.~\ref{fig-Green_T_T0} $G^+_L(\tau)/[s(\tau)]^{-g/2}$ as a function of $\tau/\beta$  for a wide range of coupling constants  $\lambda_B$ using a {\em fixed} ratio
 of $T/T^*\approx 0.014$. The perfect collapse of the data shows that temperature only enters in the combination $T/T^*$ as predicted by  Eq.~(\ref{scaling}).

\begin{figure}[t!]
\begin{center}
    \includegraphics[width=0.48\textwidth]{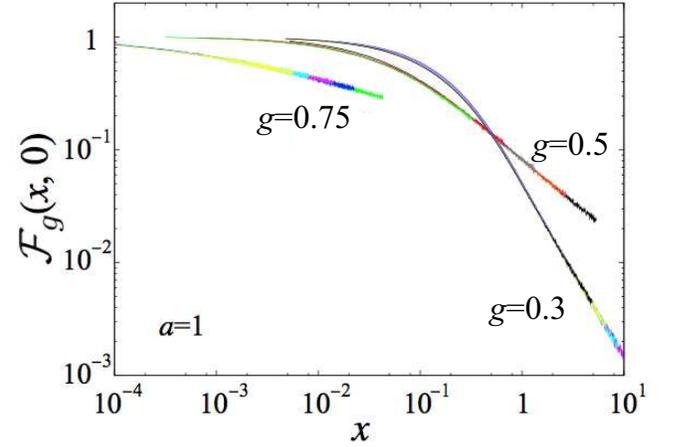}
\end{center}
\vspace{-0.5cm}
\caption{(Color online) ${\mathcal F}_g(x,0)$ vs $x$ for
 $g=0.75$, $g=0.5$, and $g=0.3$ with $a=1$. Data for $\tau/\beta<1/6$
 of each of the lines in Fig. \ref{fig-Green_scale} 
 are used in order to extract the $T=0$ limit.  
}
\label{fig-UnivF_func}
\end{figure}

Finally, we show the $T=0$ scaling functions ${\mathcal F}_g(x,0)$ for
$g=0.75$, $g=0.5$, and $g=0.3$ in Fig. \ref{fig-UnivF_func}. The $T=0$ curves can simply be obtained from the small $\tau$ data (we use $\tau/\beta<1/6$) shown in
Fig. \ref{fig-Green_scale}, which are independent of $\beta$ within the scatter of the curve.
For $g$ close to $1$, $T^*$ becomes exponentially small and it becomes
more difficult to extract the low $T$ and long $\tau$ results.

\subsection{XXZ Kondo model in helical liquids}\label{XXZ}
Along edges of two-dimensional topological insulators, a special
one-dimensional electron system is realized \cite{Wu}. Namely right- and
left-moving electrons have {\em opposite} spin polarizations, up and down, respectively. 
The topological protection of these edge channels is reflected by unusual scattering properties:
due to time-reversal symmetry a static potential cannot scatter a right-moving spin-up electron
into a left-moving spin-down electron.

The situation differs in the presence of a magnetic impurity. Using
a spin-flip process, 
a right mover can be converted in a left mover (and vice versa) due to the exchange interaction with the quantum impurity.
Therefore, it is an interesting problem to study magnetic quantum impurities at the edge of a topological insulator to investigate whether and how topological protection is affected by their presence.

In this subsection, we examine the spin-1/2 XXZ Kondo model \cite{Wu,Mac0,Mac}. We restrict our analysis to
the case where the total spin in the $z$ direction is conserved.  Although this symmetry is 
broken in real materials, e.g., by Rashba interactions \cite{Erik,Alt}, it is important to 
clarify also the basic properties of this simplified problem.  Note that the transport properties
in the presence and absence of this symmetry qualitatively differ as the current in a helical edge state (proportional to $N_\uparrow - N_{\downarrow}$) can only degrade by processes where spin conservation is violated.

We will use  different units from those in the previous subsection, and use the energy unit $v/\xi=1$ for all $g$
and $\xi=1$ as a unit of length, in order to use the same high-energy cutoff as
in  previous studies \cite{Wu,Mac0,Mac}. The main results in this
subsection are the phase diagram in $g$-$\lambda_B$ plane, which has
been discussed perturbatively \cite{Wu,Mac} and
the numerically-exact time and temperature dependence of the spin-spin correlation functions
for general interaction parameters.

\subsubsection{Model}
For XXZ Kondo model, $\hat{X}_{F}^{\sigma}$ and $\hat{X}_B$ in
Eqs. (\ref{HimpF}) and (\ref{Himp}) are given as 
\begin{eqnarray}
\hat{X}_F^{+}=\hat{S}^z,\quad
\hat{X}_F^{-}=1,\quad {\rm and}\ \ \ \hat{X}_B=\hat{S}^-.
\end{eqnarray}
We have used a slightly
different quantization axis of the impurity spin from in
 Refs.~\cite{Mac0,Mac}: $\hat{S}_z\leftrightarrow-\hat{S}_z$ and
$\hat{S}^{\pm}\leftrightarrow \hat{S}^{\mp}$.
Since  $V_F^-$ term is a pure potential scattering in the charge sector
and equivalently $\Phi_-$ sector, this does not affect the CTQMC and the
following discussions, we will concentrate on the $\Phi_+$
sector, $V_F^++V_B$, hereafter. 

First, we write the Hamiltonian in the
bosonization basis
\begin{eqnarray}
H&=&H_{1D}\nonumber\\
 &&+\lambda_F\sqrt{2/g} \partial_x \Phi_+(0) \hat{S}^z
+\tilde{\lambda}_B F_L^{\dg}F_R \hat{V}_{+\sqrt{2g}}(\Phi_+)\hat{S}^-\nonumber\\
&&+\tilde{\lambda}^*_B F_R^{\dg}F_L
 \hat{V}_{-\sqrt{2g}}(\Phi_+)\hat{S}^+, \label{modelXXZb}
\end{eqnarray}
where $\lambda_F=J_z a/2 \pi$ describes the coupling of the
$z$-component of the spin, while $\lambda_B=J_\perp a/2 \pi$
parametrizes the strength of spin-flip terms \cite{Mac0}.

For the CTQMC, it is useful to transform $H$ via a unitary
transformation 
$\hat{U}$ \cite{Mac},
\begin{equation}
\hat{U} \equiv \exp\Bigg[
 i\frac{\sqrt{2g}\lambda_F}{gv}\Phi_+(0)\hat{S}^z\Bigg]. 
\end{equation}
This erases the $\lambda_F$ term in Eq. (\ref{modelXXZb}), since
\begin{equation}
\hat{U}H_{1D}\hat{U}^{\dg}=H_{1D}-\frac{\sqrt{2g}\lambda_F}{gv}\cdot v \hat{S}^z \partial_x\Phi_+(0).
\end{equation}
Thus, the Hamiltonian is transformed to
\begin{eqnarray}
\hat{U}H\hat{U}^{\dagger}&=&H_{1D}
+{\lambda}^{\prime}_B F_L^{\dg}F_R \hat{V}_{+\lambda'}(\Phi_+)\hat{S}^-\nonumber\\
&&+{\lambda^{\prime*}_B} F_R^{\dg}F_L \hat{V}_{-\lambda'}(\Phi_+)\hat{S}^+,
 \label{HimpBXY}
\end{eqnarray}
with 
$
\lambda'=g'\sqrt{2/g}$ and $\lambda'_B=\lambda_B a^{{g'}^2/g-1}$, where
$g'$ is defined as 
\begin{eqnarray}
g'=g-\lambda_F/v. \label{gprime}
\end{eqnarray}
As will be discussed in Appendix \ref{sym}, it is sufficient to consider
cases for $\lambda_F\le gv$, i.e., $\lambda'\ge 0$.

The CTQMC algorithm for this model is similar to that for the
Kane-Fisher model. Indeed, an exact relation between the partition
functions of the two models is known \cite{FendleyZ}.
Only even $N=2k$ order terms remain finite due to the fact
that the impurity spin is 1/2, i.e., $\hat{S}^+\hat{S}^+=\hat{S}^-\hat{S}^-=0$ and/or the
total fermion number conservation. This also
restricts configuration space for the impurity spin in $Z$. We just need
to generate configurations in which $S^+$ and $S^-$ appear
alternatively: 
$\hat{S}^{\pm}(\tau_1)\hat{S}^{\mp}(\tau_2)\hat{S}^{\pm}(\tau_3)\cdots$.
Thus, we can use algorithm similar to the ``segment'' representation
used in the Anderson model, which accelerates the acceptance rate in the MC
samplings \cite{Werner}.

\subsubsection{Spin-spin correlation function}
In this subsection, we explain how to calculate the dynamical
spin-spin correlation functions. 

First, let us discuss the transverse local spin susceptibility, 
$\chi^{\perp}(\tau_{ij})\equiv
[\chi^{+-}(\tau_{ij})+\chi^{-+}(\tau_{ij})]/2$, where
$\chi^{\pm\mp}(\tau_{ij})$ is defined as 
\begin{eqnarray}
\chi^{+-}(\tau_{ij})\equiv \langle T_{\tau}
 \hat{S}^{+}(\tau_i)\hat{S}^{-}(\tau_j) \rangle.
\end{eqnarray}
Noting that $\hat{U}\hat{S}^{\pm}\hat{U}^{\dg}=e^{\pm i \sqrt{2g}\lambda_F/(gv)\Phi_+(0)}\hat{S}^{\pm}$, 
we can calculate \begin{equation}
\chi^{+-}(\tau)=\frac{1}{\beta} \left\langle \sum_{ij}^k {\mathcal
				 M}_{ij}\Big[\delta(\tau-\tau_{ij})+\delta(\beta+\tau_{ij}-\tau)\Big]\right\rangle,\
\ \ 
\end{equation} by sampling the following
quantity:
\begin{eqnarray}
{\mathcal M}_{ij}\!&=&\!
\frac{a^{2g(\frac{\lambda_F}{gv})^2}}{|\lambda^{\prime}_B|^{2}}\big[s(\tau_{ij})\big]^{-\frac{2\lambda_F}{v}}
\Bigg|\frac{
{\rm det} \hat{S}_{k-1}\{\tau\ominus \tau_i^-,\tau_j^+\}
}{
{\rm det} \hat{S}_{k}\{\tau\}
}
\Bigg|^{2g'}\nonumber \\ \label{chiperp1}
&=&
\frac{a^{2g(\frac{\lambda_F}{gv})^2}}{|\lambda^{\prime}_B|^{2}}\big[s(\tau_{ij})\big]^{-\frac{2\lambda_F}{v}}
\Big|(\hat{S}^{-1}\{\tau\})_{ji}\Big|^{2g'}, \label{chiperp2}
\end{eqnarray}
where $\tau_i$  and  $\tau_j$ are chosen in a given snapshot
$\{\tau\}$ at the $2k$th order as in Eq. (\ref{G2}) and the corresponding vertex operators at
$\tau_i$ and $\tau_j$ 
should have $\lambda_i=-\lambda'<0$ and $\lambda_j=\lambda'>0$,
respectively. 
For $\chi^{-+}(\tau_{ij})$, the same
expression holds with regarding now $\lambda_i>0$ and $\lambda_j<0$.
We also use symmetry properties
$\chi^{\pm\mp}(-|\tau|)=\chi^{\pm\mp}(\beta-|\tau|)$ to obtain results for
$0\le \tau\le \beta$. ${\mathcal M}_{ij}$ is, indeed, derived in an almost
identical way as in Appendix \ref{app-Eq2}.

Second, as $U\hat{S}^{z}U^{\dg}=\hat{S}^z$, the longitudinal spin susceptibility is directly
evaluated as
\begin{eqnarray}
&&\chi^z(\tau_{ij})= \nonumber \\
&&\frac{1}{N_{\rm MC}}\sum_{i=1}^{N_{\rm MC}}\frac{\langle \hat{S}^{\pm}(\tau_1) \cdots \hat{S}^{z}(\tau_i)
 \cdots \hat{S}^z(\tau_j) \cdots \hat{S}^{\mp}(\tau_{2k})\rangle_{\rm imp}}{
\langle \hat{S}^{\pm}(\tau_1) \cdots \hat{S}^{\mp}(\tau_{2k})\rangle_{\rm imp}
}.\ \ \ \ \ \nonumber\\
&&\label{chiZ}
\end{eqnarray}
This is possible because the operator $\hat{S}_z$ does not alter
any quantum numbers along the imaginary time axis in CTQMC. 
In contrast, the 
transverse susceptibility can be calculated only through the more complicated Eq. (\ref{chiperp2}) 
[if one would replace $\hat{S}^z(\tau_{i,j})$ by
$\hat{S}^{\pm}(\tau_{i,j})$ in Eq. (\ref{chiZ}), one would get just zero, since
the inserted $\hat{S}^{\pm}(\tau_{i,j})$ are always next to
$\hat{S}^{\pm}(\tau_{\alpha})$ with $\tau_{\alpha}\in \{\tau\}]$.

\subsubsection{SU(2) check for $g=1$ and $\lambda_F=\lambda_B$} \label{sec-checkSU2}

Interactions at the edge of the quantum wire destroy even for $\lambda_F=\lambda_B$ the SU(2) spin symmetry.
In the noninteracting electron limit ($g=1$), however, the algorithm has to recover SU(2) symmetry for $\lambda_F=\lambda_B$.
As the algorithm treats spin-flip and nonflip terms very differently,
it is a nontrivial check of the numerical data to see whether  $2\chi^z(\tau)=\chi^{\perp}(\tau)$.

\begin{figure}[t!]
\begin{center}
    \includegraphics[width=0.48\textwidth]{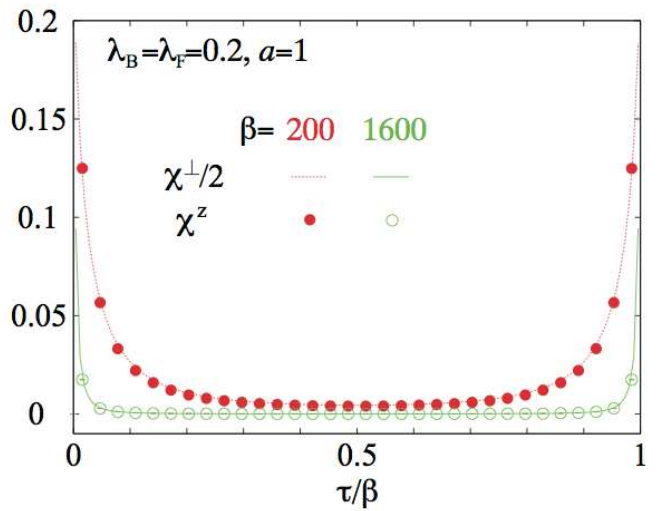}
\end{center}
\vspace{-0.5cm}
\caption{(Color online) Comparison between $\chi^{\perp}(\tau)/2$ and
 $\chi^z(\tau)$ as a function of $\tau$ for
 $\lambda_F=\lambda_B=0.2$, $a=1$, and $\beta=200$ and $1600$.}
\label{fig-SU2check}
\end{figure}

Figure \ref{fig-SU2check} shows $\chi^{z,\perp}(\tau)$ versus $\tau/\beta$ for
$\lambda_F=\lambda_B=0.2$ and $a=1$. As one can see, the relation
$2\chi^z(\tau)=\chi^{\perp}(\tau)$ holds well.  
 The intrinsic SU(2) symmetry breaking
 in the Abelian bosonization in our scheme leads to a small $\sim$10 \% deviations from
 unity for $\chi^{\perp}(\tau)/[2\chi^{z}(\tau)]$ for $a=1$. These errors do not
 alter the asymptotic behaviors for $\tau\gg 1$ and we have checked that
 the expected results $\chi^{z,\perp}(\tau)\sim \tau^{-2}$ for $g=1$ are 
 reproduced (see also Figs. \ref{fig-chip} and \ref{fig-chiz}).

\subsubsection{Phase diagram}\label{phaseKondo}

\begin{figure}[t!]
\begin{center}
    \includegraphics[width=0.48\textwidth]{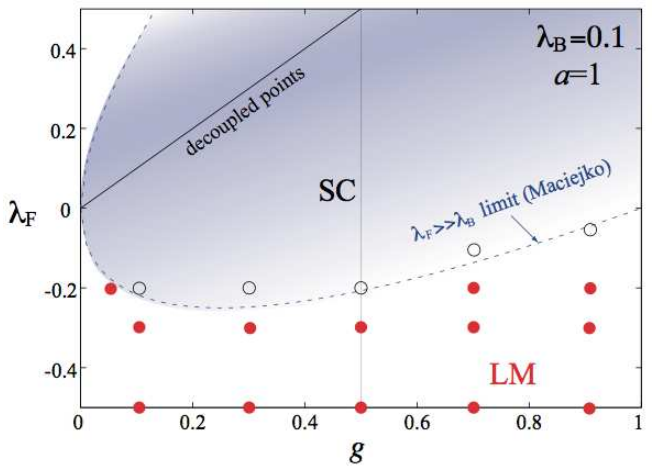}
\end{center}
\vspace{-0.5cm}
\caption{(Color online) Phase diagram as a function of the TL parameter $g$ and the size of the coupling of the $z$ component of the exchange coupling $\lambda_F$ for a fixed spin-flip rate
 $\lambda_B=0.1$  and $a=1$. Open (filled) circles indicate the screened
 (local-moment) phase. For the SC phase, points inside the phase are not
 indicated and the SC phase is symmetric with respect to the variation
 of $\lambda_F$ around the decoupled-point line (see Appendix
 \ref{sym}). The dashed line represents the phase boundary determined by
 the renormalization group analysis for $\lambda_F\gg \lambda_B$ \cite{Mac}. }
\label{fig-phase}
\end{figure}

Before starting detailed analysis, we show  in Fig. \ref{fig-phase} the global phase diagram in the plane spanned by
$\lambda_F$ and $g$ for fixed $\lambda_B=0.1$.
As pointed out in the previous studies \cite{Mac}, there are two phases: the screened phase (SC) where the Kondo effect leads to a screening of the impurity and the local moment (LM) phase where spin-flips are completely suppressed for $T \to 0$. The phase boundary for
$\lambda_B=0.1$ is well
described by the recent renormalization group results for
$\lambda_B\ll\lambda_F$ represented by the dashed line \cite{Mac}. As discussed by
Maciejko \cite{Mac} and also as discussed in Appendix \ref{sym}, the system
is symmetric at the solvable ``decoupled points'' at $\lambda_F=gv$ and the system 
for $\lambda_F>gv$ can be mapped to that for $2gv-\lambda_B<gv$, and
vice versa. Thus, we have only examined the lower part of the boundary in Fig. \ref{fig-phase}.

\begin{figure}[t!]
\begin{center}
   \includegraphics[width=0.48\textwidth]{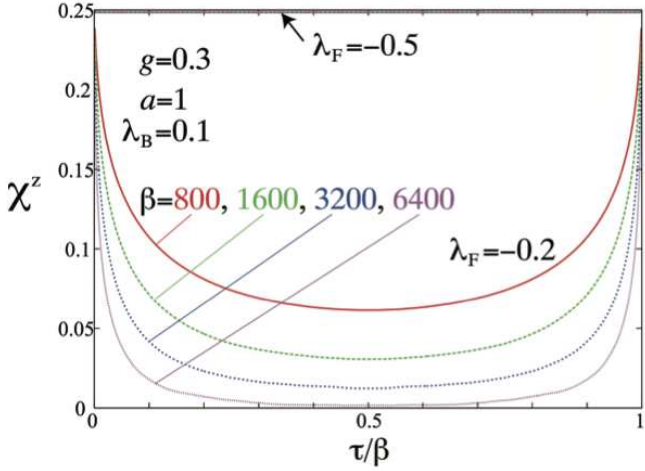}
\end{center}
\vspace{-0.5cm}
\caption{(Color online) Longitudinal dynamical local spin
 susceptibility $\chi^z(\tau)$ vs $\tau/\beta$ for
 $\lambda_F<0$ and $\lambda_B=0.1$, $a=1$, and $\beta=800$-$6400$.}
\label{fig-disting}
\end{figure}

The two phases are easily distinguished by the temperature dependence of
 $\chi^z(\tau)$ . For example, Fig. \ref{fig-disting} shows the typical behavior of the two phases
 for $g=0.3$. For the LM phase ($\lambda_F=-0.5$),
$\chi^z(\tau)$ is large and almost $\tau$-independent. Also the $T$ variations are not noticeable on the
the scale of the plot: the impurity spin is almost free and the spin-flips are strongly suppressed.
 In contrast, in the SC phase,
$\lambda_F=-0.2$, $\chi^z(\tau)$ shows a strong temperature and $\tau$ dependence,
which reflects screening processes due to the conduction electrons. In
the next subsection, we will discuss the low-temperature behaviors of
the spin-spin correlation functions in the SC phase.

\subsubsection{Dynamical local spin susceptibility}
\begin{figure}[t!]
\begin{center}
    \includegraphics[width=0.48\textwidth]{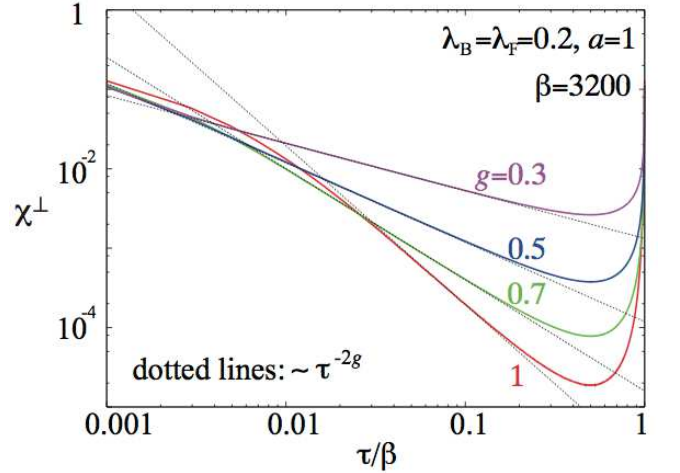}
\end{center}
\vspace{-0.5cm}
\caption{(Color online) Transverse dynamical local spin susceptibility
 $\chi^{\perp}(\tau)$ vs $\tau$ for
 $\lambda_F=\lambda_B=0.2$, $a=1$, and $\beta=3200$. The dotted lines indicate
 $\sim 1/\tau^{2g}$.}
\label{fig-chip}
\end{figure}

\begin{figure}[t!]
\begin{center}
    \includegraphics[width=0.48\textwidth]{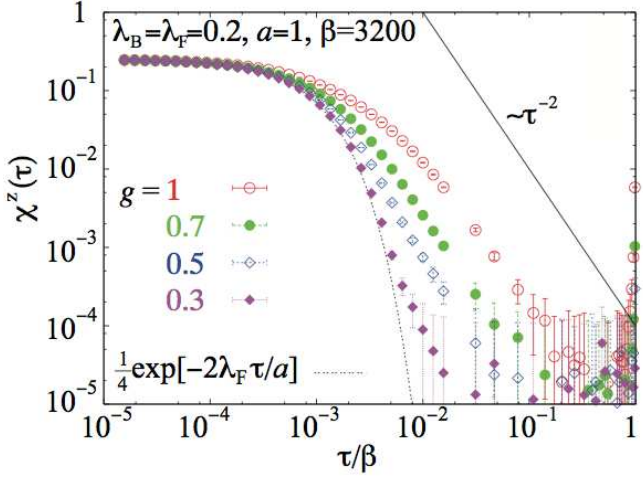}
\end{center}
\vspace{-0.5cm}
\caption{(Color online) Longitudinal dynamical local spin
 susceptibility $\chi^z(\tau)$ vs $\tau$ for
 $\lambda_F=\lambda_B=0.2$, $a=1$, and $\beta=3200$.
The dashed line shows the exact
 result for $gv=\lambda_F=0.2$: Eq. (\ref{app-chidp}),  and the
 line represents $1/\tau^2$.
}
\label{fig-chiz}
\end{figure}

\begin{figure}[t!]
\begin{center}
    \includegraphics[width=0.48\textwidth]{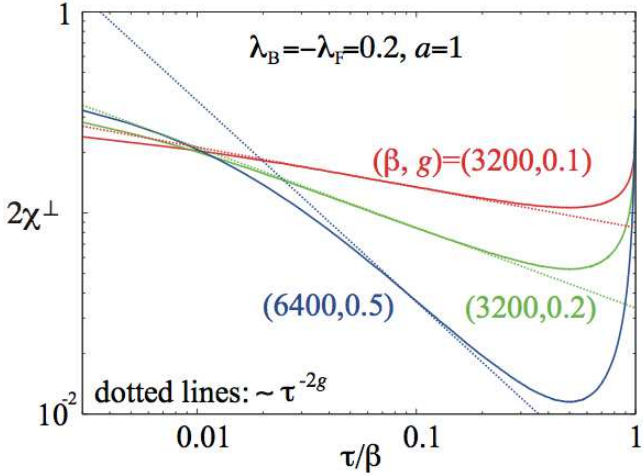}
\end{center}
\vspace{-0.5cm}
\caption{(Color online) Transverse dynamical local spin susceptibility
 $\chi^{\perp}(\tau)$ vs $\tau$ for
 $-\lambda_F=\lambda_B=0.2$, $a=1$, and $\beta=3200$ and $6400$. The
 dotted lines represent $\sim 1/\tau^{2g}$.
}
\label{fig-perpFM}
\end{figure}

Figures \ref{fig-chip} and \ref{fig-chiz} show the $\tau$ dependence of the
spin-spin correlation functions $\chi^{\perp}(\tau)$ and
$\chi^z(\tau)$, respectively, for $\lambda_{B,F}=0.2$, $\beta=3200$, and $g=0.3$, $0.5$, $0.7$, and $1$.
We find that the long-time asymptotic decay in the SC phase is given by
\begin{eqnarray}
\chi^{\perp}(\tau)\sim \tau^{-2g},
\end{eqnarray}
while
\begin{eqnarray}
\chi^{z}(\tau)\sim \tau^{-2}\quad {\rm for}\ \ g\ne \lambda_F/v.
\end{eqnarray}
These $\tau$ dependencies are  also found for $\lambda_F<0$ as long as one remains in  the SC phase
as shown in Fig. \ref{fig-perpFM}, where $\chi^{\perp}(\tau)$ are shown
for simplicity. The characteristic energy scale becomes smaller and
smaller as approaching the phase boundary (increasing $g$). For $g=0.5$, 
$\beta=6400$ is still not sufficiently low to realize complete
$\tau^{-2g}$ dependence, while for smaller $g$'s $\tau^{-2g}$ dependence
is realized already at $\beta=3200$.

Near the decoupled point at $\lambda_F=gv$, the leading power-law decay
$\tau^{-2}$ in $\chi^z(\tau)$ is suppressed and an exponential decay
appears, while for $\chi^{\perp}(\tau)$, there is no such contribution near the
decoupled point. These results
are consistent with the perturbative analysis in Appendix \ref{Pert}. 
In the following, we will concentrate on the cases for $\lambda_F\ne gv$. 
Note that the exponent of $\chi^{\perp}(\tau)$ is precisely given by that at
the decoupled point, which is related to the scaling trajectory \cite{Mac}.

These asymptotic forms readily indicate that the local spin
susceptibility $\chi^{z,\perp}(T)$ exhibits
\begin{equation}
\chi^{z,\perp}(T)=\int_0^{\beta}d\tau\chi^{z,\perp}(\tau)\sim
 T^{2\Delta_{z,\perp}-1}+{\rm const.}, \label{chi-Tdep}
\end{equation}
where the constant part comes from the short-time cutoff.
From our CTQMC results, the scaling dimensions $\Delta_{z,\perp}$ at the
screened fixed point are given by 
\begin{eqnarray}
\Delta_z=1 \quad {\rm and}\quad \Delta_{\perp}=g.
\end{eqnarray}
This is the expected result: Applying a small magnetic field to the screened magnetic impurity is equivalent
to applying a local magnetic field to the quantum wire {\em without} the
magnetic impurity. 
This problem can directly be mapped to the Kane and Fisher problem
investigated in the previous subsection. 
A magnetic field in the $z$ direction induces only forward scattering
interaction, which is not renormalized, $\Delta_z=1$, by the TLL interactions. In contrast, an
infinitesimal  
transverse magnetic field is a relevant perturbation whose 
scaling dimension $g$ can be read off from Eq. (\ref{twopoint}).

For $\Delta_{\perp}=g=1/2$, there are logarithmic corrections and 
\begin{equation}
\chi^{\perp}(T)\sim-\ln T +{\rm const.} \quad {\rm for}\ g=\frac{1}{2}. \label{chi-Tdep2}
\end{equation}
Thus, the transverse spin susceptibility for $g\le 1/2$ diverges, while
other cases and $\chi^z_s(T)$ stay constant at low temperatures.

\begin{figure}[t!]
\begin{center}
    \includegraphics[width=0.48\textwidth]{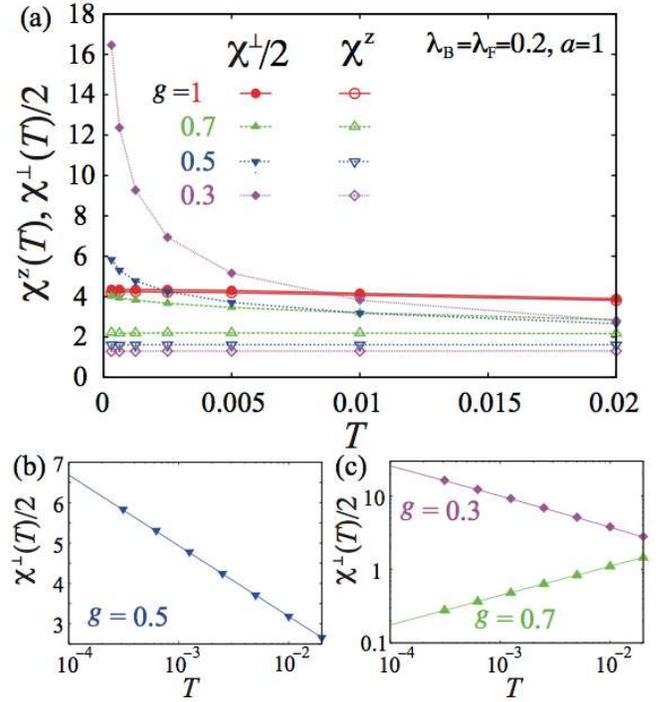}
\end{center}
\vspace{-0.5cm}
\caption{(Color online) (a) $\chi^{z}(T)$ and $\chi^{\perp}(T)/2$
 vs $T$
 for
 $\lambda_F=\lambda_B=0.2$, $a=1$, and $g=1$, $0.7$, $0.5$, and $0.3$.
(b) $\chi^{\perp}(T)/2$ vs $T$ in the log scale for $g=0.5$. The line
 indicates the fit by $-0.76\log(T/0.66)$. (c) $\chi^{\perp}(T)/2$ vs $T$
 for $g=0.7$ and $g=0.3$ in the double-log scale. For $g=0.7$,
 $4.3-\chi^{\perp}(T)/2$ is plotted and the line shows the fit by
 $7T^{0.4}$. For $g=0.3$, the line indicates the fit by $-0.396+0.66T^{-0.4}$.
}
\label{fig-chi-T}
\end{figure}

These temperature dependencies are indeed obtained from a direct numerical integration
of $\chi_{s}^{z,\perp}(\tau)$. Figure \ref{fig-chi-T} shows
$\chi^{z,\perp}(T)$ for $\lambda_{B,F}=0.2$, $a=1$, $g=1$, $0.7$,
$0.5$, and $0.3$. For $g=1$, $\chi^{\perp}(T)=2\chi^z(T)$ holds due to
the SU(2) symmetry. For other values of $g$'s, $\chi^{\perp}(T)\ne
2\chi^{z}(T)$.  Figures \ref{fig-chi-T} (b) and \ref{fig-chi-T} (c) show that the susceptibilities follow the predicted power-law
of Eq. (\ref{chi-Tdep}) with high precision for $g=0.7$ and $0.3$  and exhibit the expected logarithmic dependence for $g=0.5$, see
Eq. (\ref{chi-Tdep2}). Note that for $g=0.7$ we plot
$[\chi^{\perp}(0)-\chi^{\perp}(T)]/2$ with $\chi^{\perp}(0)/2 \approx 4.3$ in order to analyze the subleading power-law dependence.


\section{Discussions}\label{dis}
In this section, we will discuss the reliability of some expressions for correlation
functions in the ``weak-coupling'' fixed points and also discuss a
possible extension of our method to more complex problems.

\subsection{Correlation functions in the weak-coupling fixed points}\label{disWC}
For attractive interactions, $g > 1$, in the Kane-Fisher model and for the local moment phase for
the XXZ Kondo model, the nonlinear interactions are irrelevant and the
system is therefore described by a weak coupling fixed point.  In this
regime, not only do the physical properties of the model completely
differ (the impurity does not cut the chain and the local moment is not
screened)  but also 
 the statistical properties of our Monte Carlo sampling change qualitatively.
As a consequence, we find that the results based on the method defined
by Eqs. (\ref{G2}) and (\ref{chiperp1}) 
do not give reliable results, while, in contrast, the alternative approach, Eqs. (\ref{Gfin}) and (\ref{chiZ}), gives much better results.
The reason Eqs. (\ref{G2}) and (\ref{chiperp1}) are not efficient
there would be the smallness of overlap between the important
configurations 
for the partition function and those for the Green's
functions. This would be overcome by using a worm algorithm \cite{Gull}.
This is also the reason  we use
$\chi^z(\tau)$ (not affected by this problem) to identify the two phases in the XXZ Kondo
model in Sec. \ref{phaseKondo}.

\subsection{Further applications}
Here we discuss  briefly  further applications of our method  for calculating other physical quantities 
for other models. 

For impurity problems, the most important physical quantity is perhaps the conductance. It can naturally be computed within our scheme using that the 
current operator at $x=0$ is expressed by the time derivative of
$\Phi_+$ as
\begin{eqnarray}
j(x=0,\tau)=i\frac{e\sqrt{g}}{2\pi}\partial_{\tau}\Phi_+(\tau,x=0),
\end{eqnarray}
where $e$ is the elementary charge. 
The correlation function of $j(0,\tau)$ can be effectively calculated
in our CTQMC method, and, via analytic continuation to real frequency \cite{Jarrel},
one can obtain the conductance. This can be done both for the
Kane-Fisher and the XXZ Kondo problems. This approach is, however, beyond the scope of our work 
and  will be published elsewhere \cite{future}. Note that an analytic continuation to real frequencies is also needed to calculate, e.g., the tunneling density of states from our results for Green's functions.

Our method can also be directly applied to other scattering problems,
involving, for example, the backward scattering of pairs of fermions  from nonmagnetic impurities at the edge of a two-dimensional topological insulator \cite{Mac,Crepin}.
This problem is described by the same Hamiltonian as the Kane-Fisher model but, for example, the tunneling density of states, has to be computed from a different correlator.

In this paper, we have used two-component ($L$ and $R$) fermionic models
as a microscopic model for the bulk TLL. 
It is straightforward to apply this approach also to spin models or
models of bosons in all cases where these models can be described by
TLLs.
Here one can use standard bosonization identities to map those problems
to the ones considered in our paper. It is, however, important to keep
track of Klein factors in such mappings. Since our approach fully relies
on the noninteracting bosons in the bulk system, it should be
emphasized that our method cannot manage interactions of the TLL
bosons in the bulk (describing, e.g., Luther-Emery liquids \cite{LE} or
band-curvature effects).

 It would also be highly interesting to study exotic Kondo models coupled to TLLs,
which can, e.g., be realized using  Majorana modes arising from
topological edge states of 
superconducting islands \cite{Beri,Beri2,Altland} or by using two helical edges \cite{2CKHL}. With ultracold atoms, one can also realize
 Majorana edge mode
coupled to a TLL \cite{Altman}. Knowledge about the dynamics in
such problems is not accessible so far, and thus, it is interesting to
analyze them on the basis of the CTQMC developed in this paper. 

Another
technical challenge would be an analysis of impurity problems where two
relevant operators compete 
with each other (arising, e.g., for Kondo models coupled to a helical
edge when the spin in $z$ direction is not conserved) and
also investigation of
multiple and cluster impurities are highly nontrivial.  
While, for the cases discussed in this paper, no negative-sign problem
occurred, this might 
not be the case for more complex realization involving several competing scattering channels.{\vspace{0.5cm}}

\section{Summary}\label{sum}
In summary, we have demonstrated that the continuous-time quantum Monte
Carlo method can successfully be applied to situations where a quantum impurity is coupled to an {\em interacting}
one-dimensional quantum wire described by a Tomonaga-Luttinger liquid.

Our method is negative-sign free, which has been proved analytically, and
thus, very low temperature calculations are possible as demonstrated. 
The coding can be realized by a straightforward extension of existing CTQMC codes for purely fermionic problems (without interactions in the environment)
as the expression for $\delta Z_{2k}$ [Eq. (\ref{VVV-Llimit2})] are identical to those of fermionic systems 
apart from the exponent $2g$. This very simple
modification from noninteracting electron system for the bulk part
contains all the necessary information about interactions in the bulk system.

We have applied our algorithm to two models. One is the effect of a
static scattering potential in a TLL discussed by Kane and
Fisher in their classical work \cite{KF}. The second is the XXZ Kondo
model in the edge of two-dimensional topological insulators \cite{Wu,Mac0,Mac}.

For the Kane-Fisher model, we have demonstrated that the long-time
asymptotic behavior of electron Green's function 
is consistent with that predicted by Furusaki \cite{Furusaki}.
We have also computed the  universal scaling function of the Green's function for the first time.

As for the XXZ Kondo model, we have obtained the susceptibilities characterizing the two relevant fixed points: the decoupled local moment fixed point and the screened Kondo fixed point.
The temperature dependence and the asymptotic time dependence are 
consistent with analytic predictions in the whole parameter regime.

The method introduced in this paper is flexible and can be applied to 
other models and used to study transport properties. We will report 
analyses of experimentally measurable quantities via analytic
continuations and other interesting models in future publications.

\section*{Acknowledgment}
The authors thank  M. Garst and J. Kleinen  for fruitful
discussions. This work is supported by a Grant-in-Aid for Scientific
Research (Grant No. 30456199) from the Japan Society for the Promotion
of Science 
and by the
center for Quantum Matter and Materials (QM2) of the University of
Cologne. K. H. was supported by
Yamada Science Foundation for his long-term stay at the University of
Cologne. Some of the numerical calculations were done at the
Supercomputer Center at ISSP, University of Tokyo.

\appendix
\section{Electron Green's function}\label{deriG}
In this Appendix, we show detailed derivations of Eqs. (\ref{Gfin}) and 
(\ref{G2}). A similar analysis is used when we consider 
 the transverse local spin susceptibility in the XXZ Kondo model in
Sec. \ref{XXZ}. 

\subsection{Equation (\ref{Gfin})}
First, let us derive Eq. (\ref{Gfin}). We discuss one configuration 
in the sampling summation in Eq. (\ref{Av}) with $N_m=2k$. Setting 
$\hat{A}=F_L(\tau_i)\hat{V}_{-\eta}(\Phi_+,\tau_i)F_L^{\dg}(\tau_j)\hat{V}_{\eta}(\Phi_+,\tau_j)$ 
in Eq. (\ref{Av}), we obtain
\begin{eqnarray}
 {\mathcal G}_{i>j}^{(2k)}&=&\frac{\langle T_{\tau}
F_L(\tau_i)\hat{V}_{-\eta}(\tau_i) F_L^{\dg}(\tau_j)
\hat{V}_{\eta}(\tau_j)
\delta \hat{Z}_{2k}\{\tau\}\rangle_0}{\delta Z_{2k}\{\tau\}},\ \ \ \ \label{app-A1}
\end{eqnarray}
with $\tau_i>\tau_j$. Remember that $\eta=\sqrt{g/2}$ and we have abbreviated
$\hat{V}_{\eta}(\Phi_+,\tau)$ simply as $\hat{V}_{\eta}(\tau)$.

Since the Klein factors and the vertex operators commute, the two
sectors are decoupled and the former sector gives $(-1)^{P_{ij}}$ after
arranging all the Klein factors in time-ordered product and evaluating
the product, where $P_{ij}$
is the number of vertices, or equivalently the number of $\tau_{\alpha}$, between
$\tau_i$ and $\tau_j$. 

To see this, let us consider a case where $P_{ij}$ is
even. It is important to notice that there is no time-dependence in the
Klein factors for $l\to \infty$ \cite{Delft} and $(F^{\dg}_LF_R)(F^{\dg}_RF_L)=1$, since $F_{L,R}^{\dg}F_{L,R}=1$.
The Klein factors for $\tau_i>\tau_{\alpha}>\tau_j$ are rearranged
to the form
\begin{eqnarray}
(F_{L,R}^{\dg}F_{R,L}F_{L,R}^{\dg}F_{R,L})^{n_p}=(-1)^{n_p}(F_{L,R}^{\dg})^{2n_p}
 (F_{R,L})^{2n_p},\ \ \ \ \ \ 
\end{eqnarray}
with $n_p$ being an integer. Thus, the time-ordered product for
$\tau_i \ge \tau \ge \tau_j$ becomes
\begin{eqnarray}
&&F_L(\tau_i)[(-1)^{n_p}(F_{L,R}^{\dg})^{2n_p} (F_{R,L})^{2n_p}]F_L^{\dg}(\tau_j)\nonumber\\
&&=+F_LF_L^{\dg}[(-1)^{n_p}(F_{L,R}^{\dg})^{2n_p} (F_{R,L})^{2n_p}].
\end{eqnarray}
This means the factor arising after the time-ordering is $+1$ when
$P_{ij}$ is even. 

When $P_{ij}$ is an odd integer, then
the Klein factors for $\tau_i>\tau_{\alpha}>\tau_j$ are reduced to 
\begin{eqnarray}
(-1)^{n_p}(F_{L,R}^{\dg})^{2n_p} (F_{R,L})^{2n_p}
F_{L,R}^{\dg}F_{R,L}.
\end{eqnarray}
Thus, 
\begin{eqnarray}
&&F_L(\tau_i)[(-1)^{n_p}(F_{L,R}^{\dg})^{2n_p} (F_{R,L})^{2n_p}F_{L,R}^{\dg}F_{R,L}]
F_L^{\dg}(\tau_j)\nonumber\\
&&=-F_LF_L^{\dg}[(-1)^{n_p}(F_{L,R}^{\dg})^{2n_p} (F_{R,L})^{2n_p}F_{L,R}^{\dg}F_{R,L}].
\end{eqnarray}
These verify that 
 the factor after time-ordering the Klein factors are
 $(-1)^{P_{ij}}\equiv p_{ij}$.

Now, Eq. (\ref{app-A1}) becomes
\begin{eqnarray}
 {\mathcal G}_{i>j}^{(2k)}&=&{p_{ij}}\frac{
\langle \hat{V}_{\lambda_1}(\tau_1) \cdots
\hat{V}_{-\eta}(\tau_i) 
\cdots
\hat{V}_{\eta}(\tau_j) \cdots \hat{V}_{\lambda_{2k}}(\tau_{2k})
\rangle_0}
{\langle \hat{V}_{\lambda_1}(\tau_1) \cdots \hat{V}_{\lambda_{2k}}(\tau_{2k}) \rangle_0}.\nonumber\\
\label{VVoverVV}
\end{eqnarray}
Here, the product is time-ordered; $\tau_1>\tau_2>\cdots
>\tau_i>\cdots>\tau_j>\cdots >\tau_{2k}$,  
and in $\delta Z_{2k}\{\tau\}$ and 
$\delta \hat{Z}_{2k}\{\tau\}$, we have denoted each vertex
operator as $\hat{V}_{\lambda_{\alpha}}(\tau_{\alpha})$ with
$\lambda_{\alpha}=\pm \sqrt{2g}$ ($\alpha=1,2,\cdots, 2k$).
Equation (\ref{VVoverVV}) is calculated by using
Eq. (\ref{VVV}), leading to 
\begin{eqnarray}
 {\mathcal G}_{i>j}^{(2k)}&=&
{p_{ij}}\frac{\prod_{\alpha>\gamma}^{2k\oplus
  ij}\big[s(\tau_{\alpha\gamma})\big]^{\lambda_{\alpha}\lambda_{\gamma}}}{\prod_{\alpha'>\gamma'}^{2k}
\big[s(\tau_{\alpha'\gamma'})\big]^{\lambda_{\alpha'}\lambda_{\gamma'}}
}.
\end{eqnarray}
Here, in the numerator, 
if $\alpha,\gamma=i$ or $j$ in the product, $\lambda_{i}=-\lambda_j=-\eta$.
It is evident that factors $s(\tau_{\alpha\gamma})$ within $\{\tau\}$
cancel out and we obtain
\begin{eqnarray}
 {\mathcal G}_{i>j}^{(2k)}&=&{p_{ij}}
\big[s(\tau_{ij})\big]^{-\eta^2}
\prod_{\gamma}^{2k}
\big[ s(\tau_{i\gamma})\big]^{-\eta\lambda_{\gamma}}
\prod_{\alpha}^{2k}
\big[ s(\tau_{\alpha j})\big]^{\eta\lambda_{\alpha}}\nonumber\\
&=&{p_{ij}}
\big[s(\tau_{ij})\big]^{\frac{g}{2}}
\Bigg(
\frac{
\prod_{\alpha>\gamma}^{2k\oplus ij}
\big[ s(\tau_{\alpha\gamma})\big]^{w_{\alpha}w_{\gamma}}
}{
\prod_{\alpha'>\gamma'}^{2k}
\big[ s(\tau_{\alpha'\gamma'})\big]^{w_{\alpha'}w_{\gamma'}}
}
\Bigg)^{g},\label{appA1-fin2}
\end{eqnarray}
with $w_{\alpha,\beta}={\rm sgn}(\lambda_{\alpha,\beta})$. Note
that the factor $g$ comes from $\eta\lambda_{\alpha,\gamma}=\pm g$.
Finally, using the generalized Wick's theorem, we obtain Eq. (\ref{Gfin}).

\subsection{Equation (\ref{G2})}\label{app-Eq2}
Second, we will discuss Eq. (\ref{G2}). This time, the point is that we
regard a snapshot $\{\tau\}$ at $2k$th order  as  one at $2(k-1)$th
order with the remaining two $\tau_i$ and $\tau_j$ assigned to each fermion
operator for the Green's function.

Suppose that $\tau_i>\tau_j$ and the vertex operator at $\tau_i(\tau_j)$ has
$\lambda_i<0(\lambda_j>0)$ in a given snapshot $\{\tau\}$, and consider the following quantity:
\begin{eqnarray}
 {\mathcal Y}_{ij} &\equiv& \frac{(-1)^{P_{ij}}}{|\tilde{\lambda}_B|^2}
\Big(\big[s(\tau_{ij})\big]^{\lambda_i\lambda_j}\Big)^{\frac{1}{4}}
\Big(
\prod_{\gamma\ne i}^{2k} \big[s(\tau_{i\gamma})\big]^{\lambda_i\lambda_{\gamma}}
\Big)^{-\frac{1}{2}}\nonumber\\
&&\times 
\Big(
\prod_{\alpha\ne j}^{2k} \big[s(\tau_{\alpha j})\big]^{\lambda_j\lambda_{\alpha}}
\Big)^{-\frac{1}{2}}.\label{MMM}
\end{eqnarray}
When ${\mathcal Y}_{ij}$ is multiplied to $\delta Z_{2k}\{\tau\}$, we obtain
\begin{eqnarray}
 {\mathcal Y}_{ij}\delta Z_{2k}\{\tau\} &=& \frac{(-1)^{P_{ij}}}{|\tilde{\lambda}_B|^2}
\big[s(\tau_{ij})\big]^{-\eta^2}
\prod_{\gamma\ne i,j}^{2k} \big[s(\tau_{i\gamma})\big]^{-\eta\lambda_{\gamma}}
\nonumber\\
\!\!\!\!\!\!\!\!\!\!&\times& 
\prod_{\alpha\ne i,j}^{2k} \big[s(\tau_{\alpha j})\big]^{-\eta\lambda_{\alpha}}
\delta Z_{2k-2}\{\tau\ominus \tau_i,\tau_j\}\nonumber\\
\!\!\!\!\!\!\!\!\!\!&=&\langle
 F_L(\tau_i)\hat{V}_{-\eta}(\tau_i)F_L^{\dg}(\tau_j)\hat{V}_{\eta}(\tau_j)\nonumber\\
&\times& \delta\hat{Z}_{2k-2}\{\tau\ominus
 \tau_i,\tau_j\}\rangle_0.\label{correspondence}
\end{eqnarray}
Here, we have used the fact that $\eta=\sqrt{g/2}$ and
$\lambda_{\alpha}=\pm \sqrt{2g}$ with $(\alpha=1,2,\cdots,2k)$.  
Then, summing all the configurations $\{\tau\}$ and the perturbation
order leads to 
\begin{eqnarray}
&&\sum_{k,\{\tau\}} {\mathcal Y}_{ij}\delta Z_{2k}\{\tau\}=\nonumber\\
&&\sum_{k,\{\tau\}}
\frac{
\langle
 F_L(\tau_i)\hat{V}_{-\eta}(\tau_i)F_L^{\dg}(\tau_j)\hat{V}_{\eta}(\tau_j)
\delta\hat{Z}_{2k-2}\{\tau\ominus
 \tau_i,\tau_j\}\rangle_0}{\delta Z_{2k-2}\{\tau\ominus
 \tau_i,\tau_j\}}\nonumber\\
&&\times \delta Z_{2k-2}\{\tau\ominus \tau_i,\tau_j\}. \label{appA2-finEq}
\end{eqnarray}
This indicates that the sampling of ${\mathcal Y}_{ij}$ is indeed equivalent to
that of the electron Green's function $G_L^{+}(\tau_{ij})$. A similar
transformation to those used in Eq. (\ref{appA1-fin2}) 
and the generalized Wick's theorem
can be applied
to Eq. (\ref{MMM}) to
 obtain Eq. (\ref{G2}), where ${\mathcal Y}_{ij}=\tilde{\mathcal G}_{i>j}^{(2k)}$.

\section{Parameter space of the XXZ Kondo model}\label{sym}
In this appendix, we briefly discuss that, for the XXZ Kondo model,
a system with $\lambda_F^{(1)}>g v$ is equivalent to a model with $\lambda_F^{(2)}=2 g v-\lambda_F^{(1)}$.
For example, a very large antiferromagnetic
$\lambda_F$ reduces to a large ferromagnetic $\lambda_F<0$ in
the transformed system. Physically, this happens by binding electrons to the impurity spin.
The symmetric point $\lambda_F=gv$ is indeed a solvable point of the
present model because $\lambda'=0$ in Eq. (\ref{HimpBXY}). 
This equivalence is understood as follows. 
For $\lambda_F>gv$, the Hamiltonian is given as
\begin{eqnarray}
UHU^{\dagger}&=&H_{1D}
+\lambda^{\prime}_B F^{\dg}_LF_R
 \hat{V}_{-|\lambda'|}(\Phi_+)\hat{S}^-\nonumber\\
&&+ \lambda^{\prime*}_BF^{\dg}_RF_L
 \hat{V}_{|\lambda'|}(\Phi_+)\hat{S}^+.\label{HimpBXY2}
\end{eqnarray}
We now interchange the up- and the down-spins for the local moment. Then, 
since the Klein factors do not matter at all by an appropriate relabeling, 
the resultant form is equivalent to Eq. (\ref{HimpBXY}),    
if $|\lambda'|$ in Eq. (\ref{HimpBXY2})
is equal to
$\lambda'$ in Eq. (\ref{HimpBXY}); $\lambda_F^{(1)}/(gv)-1=1-\lambda_F^{(2)}/(gv)$, 
 with $\lambda_F^{(1)}>gv$ and $\lambda_F^{(2)}<gv$, leading to 
$\lambda_F^{(1)}/(gv)=2-\lambda_F^{(2)}/(gv)$. This symmetry was first
taken into account in a recent renormalization group analysis \cite{Mac}.

\section{Spin-spin correlations around decoupled points}\label{Pert}
In this appendix, we first review the results for decoupled points at
$\lambda_F=gv$ in
the XXZ Kondo model discussed in Ref.~\cite{Mac}. Then we will discuss
the effects of deviations from $\lambda_F=gv$. 
\subsubsection{Decoupled points}
In this subsection, we summarize the results of the local spin
susceptibilities at decoupled points \cite{Mac}. 

For the decoupled points, $\lambda_F=gv$, the Hamiltonian reads
\begin{eqnarray}
\hat{U}H_{\rm dp}\hat{U}^{\dag}&=&H_{1D}+\frac{\lambda_B}{a}\Big[F_L^{\dag}F_R \hat{S}^- + {\rm H.c.}\Big].
\end{eqnarray}
Since the Klein factors do nothing in the following discussions about
the spin susceptibilities, this is equivalent to 
\begin{eqnarray}
\hat{U}H_{\rm dp}\hat{U}^{\dag}&=&H_{1D}+h( \hat{S}^++\hat{S}^-),
\end{eqnarray}
which is just the single-spin Hamiltonian under the magnetic field $h$ parallel
to the $x$ direction with $h=\lambda_B/a>0$ and the bosons and the spin 
are decoupled. Thus, for any values of $\lambda_B$, this can be easily
diagonalized.

We now take a new quantization axis parallel to the original $x$
direction, and then
\begin{eqnarray}
\hat{S}^{\pm}=\tilde{S}^z\mp \frac{1}{2}(\tilde{S}^+-\tilde{S}^-), \quad
\hat{S}^z=\frac{1}{2}(\tilde{S}^++\tilde{S}^-).
\end{eqnarray}
Let us list correlation functions of $\tilde{S}$ as follows:
\begin{eqnarray}
\tilde{\chi}_{+-}(\tau)&=&\langle T_{\tau}
 \tilde{S}^+(\tau)\tilde{S}^-(0)\rangle=\frac{e^{-2(\beta-\tau)h}}{1+e^{-2\beta h}},\label{app-B4}\\
\tilde{\chi}_{-+}(\tau)&=&\langle T_{\tau}
 \tilde{S}^-(\tau)\tilde{S}^+(0)\rangle=\frac{e^{-2h\tau}}{1+e^{-2\beta h}},\label{app-B5}\\
\tilde{\chi}_{zz}(\tau)&=&\langle T_{\tau}
 \tilde{S}^z(\tau)\tilde{S}^z(0)\rangle=\frac{1}{4}.\label{app-B6}
\end{eqnarray}
The local spin susceptibilities for $\hat{S}$'s are in linear
combinations of Eqs. (\ref{app-B4})-(\ref{app-B6}). Thus, we obtain for
$T=0$ 
\begin{eqnarray}
{\chi}^{\rm dp}_{+-}(\tau)&=&\langle T_{\tau}
 \hat{S}^+(\tau)\hat{S}^-(0)\rangle=\frac{1+e^{-2h\tau}}{4} \Big(
 \frac{a}{v\tau}\Big)^{2g},\ \ \ \label{chidpPM0}\\
{\chi}^{\rm dp}_{zz}(\tau)&=&\langle T_{\tau}
 \hat{S}^z(\tau)\hat{S}^z(0)\rangle=\frac{1}{4}e^{-2h\tau}.\label{app-chidp}
\end{eqnarray}
Here, we have used $\hat{U}\hat{S}^{\pm}\hat{U}^{\dag}=e^{\pm \sqrt{2g}\Phi_+}\hat{S}^{\pm}$ for
$\lambda_F=gv$.

\subsubsection{perturbations}
Let us consider the cases where $\lambda_F$ slightly deviates from $gv$:
$\delta\lambda_F=\lambda_F-gv$. Then there appears a coupling between
the bosons and the local spin as
\begin{eqnarray}
\hat{U}\delta H\hat{U}^{\dag} &=& \delta \lambda_F
 \sqrt{\frac{2}{g}}\partial_x\Phi_+(0) \hat{S}^z\nonumber\\
&=&\delta \lambda_F \sqrt{\frac{1}{2g}}\partial_x \Phi_+(0)(\tilde{S}^++\tilde{S}^-).
\end{eqnarray}
One can calculate the corrections to $\chi_{zz}^{\rm dp}$ in the
perturbation theory. 
The second-order perturbation gives
\begin{eqnarray}
\delta\chi_{zz}^{\rm dp}(\tau)=\frac{[{\rm Tr} e^{-\beta H_{\rm dp}}\hat{S}^z(\tau)\hat{S}^z(0)]^{(2)}}{Z_{\rm
 dp}Z_{\Phi_+}^0}-\frac{Z^{(2)}}{Z_{\rm dp}Z^0_{\Phi_+}}\chi_{zz}^{\rm dp}(\tau),\nonumber\\\label{dchiDP}
\end{eqnarray}
where the trace is taken over both the local spin and the $\Phi_+$ boson
parts and the superscript $^{(2)}$ indicates the second-order contribution. 
$Z_{\Phi_+}^0$ is the partition function of free $\Phi_+$ sector and $Z_{\rm dp}=e^{\beta h}+e^{-\beta h}$. 
The time dependence that differs from $\chi_{zz}^{\rm dp}(\tau)$ comes
from the first term. At $T=0$, we find that the power-law dependence appears from 
\begin{eqnarray}
&&\frac{[{\rm Tr} e^{-\beta H_{\rm dp}}{\tilde{S}}^-(\tau){\tilde{S}}^+(0)]^{(2)}}{4Z_{\rm
 dp}Z_{\Phi_+}^0}
\simeq\frac{\delta\lambda_F^2}{8gZ_{\rm dp}}\int_0^{\tau}
\!\!\!d\tau_1\int_0^{\tau_1}\!\!\!\!d\tau_2 \nonumber\\
&&\ \ \ \ \ \ \ \ \times\frac{
[{\rm Tr} e^{-2\beta h\tilde{S}^z}
 \tilde{S}^-(\tau)\tilde{S}^+(\tau_1)\tilde{S}^-(\tau_2)\tilde{S}^+(0)]
}
{v^2(\tau_1-\tau_2+a/v)^2}\label{rhs1}\\
&&\ \ \ \ \ \ \equiv
\frac{\delta\lambda_F^2}{8gv^2}I_{1}(2h\tau), \label{app-B12}
\end{eqnarray}
and 
\begin{eqnarray}
&&\frac{[{\rm Tr} e^{-\beta H_{\rm dp}}{\tilde{S}}^+(\tau){\tilde{S}}^+(0)]^{(2)}}{4Z_{\rm
 dp}Z_{\Phi_+}^0}
=\frac{\delta\lambda_F^2}{8gZ_{\rm dp}}\int_{\tau}^{\beta\to \infty}
\!\!\!\!\!\!\!\!d\tau_1\int_0^{\tau}d\tau_2 \nonumber\\
&&\ \ \ \ \ \ \ \ \times
\frac{
[{\rm Tr} e^{-2 \beta h \tilde{S}^z}
\tilde{S}^-(\tau_1)\tilde{S}^+(\tau)\tilde{S}^-(\tau_2)\tilde{S}^+(0)]
}{v^2(\tau_1-\tau_2+a/v)^2}\label{rhs2}\\
&& \ \ \ \ \ \ \equiv 
\frac{\delta\lambda_F^2}{8gv^2}I_{2}(2h\tau). \label{app-B13}
\end{eqnarray}
Here, in the right-hand side of Eqs. (\ref{rhs1}) and (\ref{rhs2}), the trace
is over the local spin configuration. 
In the right-hand side of Eq. (\ref{rhs1}), we have retained dominant terms
for large $\tau$ and ignored a diverging term 
for $T\to 0$ that cancels
out with the second term in Eq. (\ref{dchiDP}). 
Note that only terms with $\tilde{S}^+(0)$ are relevant, since a state
with $\tilde{S}^z=\downarrow$ is the ground state at the decoupled point.
Parameterizing $t=2h\tau$, $c=2ha/v$, and $b=2h\beta\to \infty$, we
obtain 
\begin{eqnarray}
I_{1}(t)&=& e^{-t}
\int_0^{t}dx \int_0^x \!dy \frac{e^{x-y}}{(x-y+c)^2}
\simeq
\frac{1}{t^2}+\cdots,\ \ \ \ \ \ \\
I_{2}(t)&=&e^{t}\int_{t}^{b}dx \int_0^t dy
 \frac{e^{-x-y}}{(x-y+c)^2}\simeq \frac{1}{t^2}+\cdots.
\end{eqnarray}
Thus, finally, we obtain
\begin{eqnarray}
\delta\chi_{zz}^{\rm dp}(\tau)\simeq \frac{\delta\lambda_F^2}{16g \lambda_B^2}\left(\frac{a}{v\tau}\right)^2.
\end{eqnarray}
This indicates that the exponential decay at the decoupled point 
immediately disappears and the leading term becomes ``Fermi liquid'' like $\sim
\tau^{-2}$.

As for the corrections $\delta\chi_{+-}^{\rm dp}(\tau)$, there appears at
least a factor $\langle
\hat{V}_{\sqrt{2g}}(\tau)\hat{V}_{-\sqrt{2g}}(0)\rangle\propto \tau^{-2g}$.
Thus, the leading $\tau$ dependence of $\chi_{+-}(\tau)$ for $\tau\to \infty$ does not change from Eq. (\ref{chidpPM0}).


\vspace{-0.5cm}

\begin{thebibliography}{99} 
\bibitem{TLL} S. Tomonaga, Prog. Theor. Phys. {\bf 5}, 544 (1950),
	J. M. Luttinger, J. Math. Phys. {\bf 4}, 1154 (1963).
\bibitem{Giam} T. Giamarchi, {\it Quantum Physics in One Dimension}, (Oxford University Press, Oxford, 2004).
\bibitem{KF} C. L. Kane and M. P. A. Fisher, Phys. Rev. Lett. {\bf 68},
	1220 (1992); Phys. Rev. B {\bf 46}, 15233 (1992).
\bibitem{Delft} J. v. Delft and H. Schoeller, Ann. Pnys. (Leipzig) {\bf
	7}, 225 (1998).
\bibitem{Furusaki} A. Furusaki, Phys. Rev. B {\bf 56}, 9352 (1997).
\bibitem{Bethe} P. Fendley, H. Saleur, N. P. Wamer, Nucl. Phys. B{\bf
	430}, 577 (1994), 
P. Fendley, A. W. W. Ludwig, and H. Saleur, Phys. Rev.
Lett. {\bf 74}, 3005 (1995), 
P. Fendley, F. Lesage, and H. Saleur, J. Stat. Phys. {\bf 85}, 211 (1996).
\bibitem{fRG} T. Enss, V. Meden, S. Andergassen,
	X. Barnab\'{e}-Th\'{e}riault, W. Metzner, and
	K. Sch\"{o}nhammer, Phys. Rev. B {\bf 71}, 155401 (2005).
\bibitem{PIMC1} K. Moon, H. Yi, C. L. Kane, S. M. Girvin, and
	M. P. A. Fisher, Phys. Rev. Lett. {\bf 71}, 4381 (1993).
\bibitem{PIMC2}	K. Leung, R. Egger, and C. H. Mak, Phys. Rev. Lett. {\bf 75},
	3344 (1995).
\bibitem{PIMC3} Y. Hamamoto, K. I.  Imura, and T. Kato, Phys. Rev. B {\bf
	77}, 165402 (2008).
\bibitem{nrg} A. Freyn and S. Florens, Phys. Rev. Lett. {\bf 107},
	017201 (2011).
\bibitem{Rubtsov} A. N. Rubtsov, V. V. Savkin, and A. I. Lichtenstein,
	Phys. Rev. B {\bf 72}, 035122 (2005). 
\bibitem{Werner} P. Werner, A. Comanac, L. de' Medici, M. Troyer, and A. J.
	Millis, Phys. Rev. Lett. {\bf 97},
	076405 (2006).
\bibitem{Otsuki} J. Otsuki, H. Kusunose, P. Werner, and Y. Kuramoto,
	J. Phys. Soc. Jpn {\bf 76}, 114707 (2007).
\bibitem{Gull} For review of CTQMC, see, E. Gull, A. I. Lichtenstein,
	A. N. Rubtsov, M. Troyer, and P. Werner,  Rev. Mod. Phys. {\bf
	83}, 349 (2011). 
\bibitem{DMFT} A. Georges, G. Kotliar,W. Krauth, and M. J. Rozenberg, Rev. Mod.
Phys. {\bf 68}, 13 (1996).
\bibitem{Anders} P. Anders, E. Gull, L. Pollet, M. Troyer, and
	P. Werner, New J. Phys. {\bf 13}, 075013 (2011).
\bibitem{Otsuki-b} J. Otsuki, Phys. Rev. B {\bf 87}, 125102 (2013).
\bibitem{Wu} C. Wu, B. A. Bernevig, and S. C. Zhang,
	Phys. Rev. Lett. {\bf 96}, 106401 (2006).
\bibitem{Mac0} J. Maciejko, C. Liu, Y. Oreg, X.-L. Qi, C. Wu, and
	S.-C. Zhang, Phys. Rev. Lett. {\bf 102}, 256803 (2009).
\bibitem{Mac} J. Maciejko, Phys. Rev. B {\bf 85}, 245108 (2012).
\bibitem{Jarrel} M. Jarrell and J. E. Gubernatis: Phys. Rep. {\bf 269}, 133 (1996).
\bibitem{Erik} E. Eriksson, A. Str\"{o}m, G. Sharma, and H. Johannesson,
	Phys. Rev. B {\bf 86}, 161103(R) (2012).
\bibitem{Alt} B. L. Altshuler, I. L. Aleiner, and V. I. Yudson,
	Phys. Rev. Lett. {\bf 111}, 086401 (2013).
\bibitem{FendleyZ} P. Fendley and H. Saleur, Phys. Rev. Lett. {\bf 75},
	4492 (1995).
\bibitem{future} K. Hattori and A. Rosch, unpublished.
\bibitem{Crepin} F. Cr\'epin, J. C. Budich, F. Dolcini, P. Recher, and
	B. Trauzettel, Phys. Rev. B {\bf 86}, 121106(R) (2012).
\bibitem{LE} A. Luther and V. J. Emery, Phys. Rev. Lett. {\bf 33}, 589 (1974).
\bibitem{Beri} B. B\'{e}ri and N. R. Cooper, Phys. Rev. Lett. {\bf 109},
	156803 (2012).
\bibitem{Beri2} B. B\'{e}ri, Phys. Rev. Lett. {\bf 110}, 216803 (2013).
\bibitem{Altland} A. Altland and R. Egger, Phys. Rev. Lett. {\bf 110},
	196401 (2013).
\bibitem{2CKHL} T. Posske, C. X. Liu, J. C. Budich, and B. Trauzettel,
	Phys. Rev. Lett. {\bf 110}, 016602 (2013).
\bibitem{Altman} J. Ruhman and E. Altman, arXiv:1401.7343.
\end{thebibliography}
\end{document}